\documentclass[aps,prx,reprint, superscriptaddress,floatfix,export]{revtex4-2} 
\usepackage[colorlinks=true,linkcolor=blue,citecolor=blue,urlcolor=blue]{hyperref}
\hypersetup{
    pdftitle={NISQSPTMBQC},
}

\usepackage[utf8]{inputenc}
\usepackage[dvipsnames]{xcolor}
\usepackage[titletoc,toc,title]{appendix}
\usepackage{times}
\usepackage{tikz}
\usepackage{quantikz}
\usepackage{amsthm,physics,empheq,tcolorbox,enumerate,comment}
\usepackage{float}
\usetikzlibrary{shapes.geometric, arrows, shapes.multipart}
\usepackage{booktabs, tabularx}

\usepackage[export]{adjustbox}
\usepackage{titlesec,amsfonts,braket,csquotes}
\usepackage{amssymb}

\def\<#1>{\mathinner{\langle#1\rangle}}

\begin{document}

\title{Testing measurement-based computational phases of quantum matter on a quantum  processor}

\date{\today}

\author{Ryohei Weil}
\email{ryoheiweil@uchicago.edu}
\affiliation{Department of Physics, University of Chicago, USA}
\affiliation{Pritzker School of Molecular Engineering, University of Chicago, USA}

\author{Dmytro Bondarenko}
\affiliation{Department of Physics and Astronomy,
University of British Columbia, Vancouver, Canada}
\affiliation{Stewart Blusson Quantum Matter Institute, University of British Columbia, Vancouver, Canada}

\author{Arnab Adhikary}
\affiliation{Department of Physics and Astronomy,
University of British Columbia, Vancouver, Canada}
\affiliation{Stewart Blusson Quantum Matter Institute, University of British Columbia, Vancouver, Canada}
\affiliation{Institut f\"{u}r Theoretische Physik, Leibniz Universit\"{a}t Hannover, Hannover, Germany}

\author{Robert Raussendorf}

\affiliation{Institut f\"{u}r Theoretische Physik, Leibniz Universit\"{a}t Hannover, Hannover, Germany}
\affiliation{Stewart Blusson Quantum Matter Institute, University of British Columbia, Vancouver, Canada}

\begin{abstract}

Many symmetry protected or symmetry enriched phases of quantum matter have the property that every ground state in a given such phase endows measurement based quantum computation with the same computational power. Such phases are called computational phases of quantum matter. Here, we  experimentally verify four theoretical predictions for them on an IBM superconducting quantum device.  We comprehensively investigate how symmetric imperfections of the resource states translate into logical decoherence, and how this decoherence is mitigated.
In particular, the central experiment probes the scaling law from which the uniformity of computational power follows. We also analyze the correlated regime, where local measurements give rise to logical operations collectively. We test the prediction that densest packing of a measurement-based algorithms remains the most efficient, in spite of the correlations. Our experiments corroborate the operational stability of measurement based quantum computation in quantum phases of matter with symmetry. 
\end{abstract}

\maketitle

\section{Introduction}\label{sec:intro}
Computational phases of quantum matter  \cite{doherty_identifying_2009, else_symmetry-protected_2012, else_symmetry_2012, miller_resource_2015, miyake_quantumcompedge_2010} describe the uniform quantum computational usefulness of states in certain physical phases of matter with symmetry, such as symmetry protected and symmetry enriched topological phases. The power of such states is harnessed through measurement-based quantum computation (MBQC) \cite{raussendorf_one-way_2001},  a scheme of universal quantum computation where the process of computation is driven by measurement. No unitary evolution takes place in it. The measurements are all local, and they are applied to a suitably entangled initial state, whose entanglement is consumed in course of the computation. 

In this model, the computational power rests with the initial state, which is therefore a computational resource. Some resource states enable universal quantum computation, such as cluster states \cite{raussendorf_persistent_2001} in spatial dimension 2 and higher, and Affleck-Kennedy-Lieb-Tasaki states \cite{aklt_rigorous_1987} of spin 3/2 \cite{miyake_quantumcapability_2011, wei_aklt_2011, darmawan_measurement-based_2012} and 2 \cite{wei_universal_2015}. 

Moving beyond examples for resource states, it is known that the overwhelming majority of quantum states---as measured by Hilbert space volume---are not useful at all for measurement-based quantum computation \cite{gross_most_2009, bremner_randompure_2009}. They are simply `too entangled to be computationally useful' \cite{gross_most_2009}. 

The existence of universal resource states on one hand and their scarcity in Hilbert space on the other make it desirable to establish a classification for them.
To date, such a classification has remained elusive. 

Yet, one more general fact is known about MBQC resource states: they are not isolated points in Hilbert space, but instead  form continuous manifolds---so-called {\em{computational phases of quantum matter}} \cite{doherty_identifying_2009, else_symmetry-protected_2012, else_symmetry_2012, miller_resource_2015, miyake_quantumcompedge_2010, stephen_subsystem_2019, Raussendorf2019ComputationalPhase}. These phases are, as far as is currently known, symmetry protected topologically (SPT) ordered \cite{gu_tensor-entanglement-filtering_2009, schuch_classifying_2011, chen_local_2010, chen_sptgroupcohom_2013, ogata_classification_2021} or symmetry enriched topologically (SET) ordered \cite{essin_classifying_2013, mesaros_classification_2013, lu_classification_2016, barkeshli_symmetry_2019}; see \cite{else_symmetry-protected_2012, else_symmetry_2012, miller_resource_2015, miyake_quantumcompedge_2010, stephen_subsystem_2019, Raussendorf2019ComputationalPhase,Daniel2020Archimedean,Stephen2019Subsystem, Devakul2018FractalSPT} and~\cite{Herringer2025MBQC_SET}, respectively. Both types are formed by ground states of local Hamiltonians, i.e. exist at zero temperature, and they require the presence of a suitable symmetry. What makes these phases interesting from the viewpoint of quantum computation is that their computational power for MBQC is {\em{uniform}}. That is, every state in a given such phase can be used to perform the exact same quantum computations.

The simplest scenario in which the phenomenology of interest can be studied is the 1D cluster Hamiltonian perturbed by a transversal magnetic field,
\begin{equation}\label{eq:interpolationH}
    H(\alpha) = -\cos(\alpha)\sum_{i=1}^n K_i - \sin(\alpha)\sum_{i=2}^{n-1}X_i
\end{equation}
where $K_i = Z_{i-1}X_iZ_{i+1}$ for $i = 2, \ldots, n-1$, $K_1 = X_1Z_2$ for $i = 1$ and $K_n = Z_{n-1}X_n$ for $i = n$ (chosen to be odd) are the cluster state stabilizer generators centered at site $i$. At $\alpha = 0$, the ground state is simply the $n$-qubit cluster state $\ket{C_n}$, and at $\alpha = \pi/2$ the ground state subspace is spanned by the 4-fold degenerate product states $\ket{\pm}\ket{+}\ldots \ket{+}\ket{\pm}$. The Hamiltonian of Eq.~\eqref{eq:interpolationH} possesses the symmetry:
\begin{equation}\label{eq:Z2Z2}
    \mathbb{Z}_2 \times \mathbb{Z}_2 \cong \langle Z_1X_2I_3\, ..\, X_{n-1}Z_n, X_1I_2X_3\, ..\,  I_{n-1}X_n\rangle
\end{equation}
and as such its ground states form a one-parameter family in the phase-diagram of $\mathbb{Z}_2 \times \mathbb{Z}_2$-symmetric states. In the thermodynamic limit, ground states for $0 \leq \alpha < \pi/4$ occupy the SPT phase of the cluster and have the equivalent computational power to the cluster state \cite{raussendorf_symmetry-protected_2017}. The ground states for $\pi/4 < \alpha \leq \pi/2$ are trivial and do not possess computational power. This transition is marked by the value of the string order parameters:
\begin{equation}\label{eq:SOPs}
    \sigma_{k} = \begin{cases}
         \langle I\, ..\, I Z_kX_{k+1}I_{k+2}\ldots X_{n-2}I_{n-1}X_n \rangle & \text{$k$ even}
         \\ \langle I\, ..\, I Z_kX_{k+1}I_{k+2}\ldots I_{n-2}X_{n-1}Z_n \rangle & \text{$k$ odd}
    \end{cases}
\end{equation}
dropping abruptly to zero 
at $\alpha = \pi/4$. In the bulk, faraway from the boundaries, the value of the string order parameter becomes independent of the site $k$ at which it is anchored, and we simply denote it by $\sigma$. As we describe below, the string order parameter also has computational significance: the larger $\sigma$ the more efficient the computation.

For finite chains, the decay of the string-order parameter smoothens as the notion of a phase becomes ill-defined. This not withstanding, the order parameter remains defined and its connection to computational efficiency persists \cite{raussendorf_measurement-based_2023}. The ground states of the Hamiltonian $H(\alpha)$ still interpolate between computationally useful and useless, making them an ideal testbed for our predictions.\smallskip

Before outlining the present paper, we provide a short summary on previous experiments in measurement-based quantum computation. MBQC was first experimentally demonstrated in photonic systems \cite{walther_experimental_2005}, creating a 4-qubit cluster state and performing a Grover search on a 4-item data base. Cluster states have been created on a wide range of platforms, such as in cold atoms in optical lattices \cite{mandel_controlled_2003}, in ion traps \cite{lanyon_measurement-based_2013} and photons (DV) \cite{lu_experimental_2007}, (CV) \cite{yokoyama_ultra-large-scale_2013} and coupled to quantum dots \cite{istrati_sequential_2020, ferreira_deterministic_2024}. Proof-of-principle topological error correction with 3D cluster states \cite{raussendorf_topological_2007} has also been realized \cite{yao_experimental_2012}.

In the context of computational phases of matter, earlier works have tested the uniform viability of quantum wire, i.e. transmission of quantum information without computation \cite{azses_identification_2020,jiang_generation_2025}, which is a precondition for computational phases. Also, quantum phase transitions were studied in NISQ devices, through measurement of string order parameters \cite{smith_crossing_2022}. A large body of work exists in the related field of measurement-driven quantum phase transitions, both theoretical~\cite{Li2018,Skinner2019,Chan2019,Bao2020,Iaconis2020} and experimental~\cite{Koh2022,Noel2022}.
\medskip

In this paper, we probe the viability of computational phases of quantum matter. We perform four experiments to test uniformity of computational power across a symmetry-protected phase, the $\mathbb{Z}_2 \times \mathbb{Z}_2$-symmetric 1D cluster phase.

Experiment~\#1 investigates the logical error in gate operation as a function of the symmetry-breaking measurement angle $\beta$. Theory predicts that the logical error is zero for $\beta =0 \mod \pi$ (the uniform wire case \cite{else_symmetry-protected_2012}), and is largest for odd integer multiples of $\pi/2$. We test the theory prediction for the full angular dependence of the logical error.

Experiment~\#2 relates computational efficiency to the strength of physical order. Computational efficiency is measured by a computational order parameter $\nu$ that describes the strength of the response in terms of a non-trivial logical gate to a measurement at angle $\beta$. In the 1D symmetry-protected phases such as the cluster phase, the strength of physical order is measured by a string order parameter $\sigma$. The prediction~\cite{raussendorf_measurement-based_2023} is that computational and physical order parameter agree, $\nu=\sigma$. This prediction is tested here.

Experiment \#3 is the central one. It tests the technique of {\em{splitting}}, by which the logical decoherence in computational phase of matter is overcome. Splitting of rotations is the basis for uniform computational power of MBQC across physical phases~\cite{raussendorf_symmetry-protected_2017}.

Experiment~\#4 is a refinement of Experiment~\#3, testing splitting in the so-called correlated regime, namely when the algorithmically non-trivial measurements are so close together that they do not individually, but jointly, give rise to logical operations. The prediction ~\cite{adhikary_counter-intuitive_2023} is that densest packing of algorithmically non-trivial measurements remains optimal, despite the correlations.\medskip

The remainder of this paper is organized as follows. In Section \hyperref[sec:discussion]{II} we give an overview of SPT-MBQC, and discuss the manifestation and management of decoherence within this formalism. In Section \hyperref[sec:exp0]{III}, we discuss an ``Experiment \#0'' with the aim of experimental preparation of MBQC resource states of interest. In Section \hyperref[sec:exp1]{IV}-\hyperref[sec:exp4]{VII} we discuss Experiments \#1-\#4, providing detailed motivation, experimental background, and results for each. In Section \hyperref[sec:discussion]{VIII} we conclude and discuss possibilities for future experiments that could verify the universality of computational phases.

\section{ Background}\label{sec:background}

In this section, we provide the essential background for measurement-based quantum computation on SPT-ordered states. We assume the reader is familiar with MBQC itself, and the representation of MBQC in the framework of matrix product states (MPS) \cite{gross_novel_2007}. 

{\em{Overview.}} While resource states in the same SPT phase can be used to run the same measurement-based quantum computations, they don't permit this with the same efficiency. Rather, the required computational cost increases towards the phase boundary. This is due to a phenomenon called {\em{logical decoherence}} \cite{else_symmetry-protected_2012}. It is present even if all relevant properties of the resource state are precisely known and the measurements driving the SPT-MBQC are perfect. It is caused by particularities of the resource states within a given SPT phase, specifically uncontrolled residual entanglement on top of symmetry-protected entanglement \cite{else_hidden_2013}. Logical decoherence is the main complication for SPT-ordered states as computational resources, but it can be counteracted \cite{miller_resource_2015, raussendorf_symmetry-protected_2017, stephen_computational_2017}. \smallskip

To date, there are three formalisms for reasoning about computational phases of quantum matter; the original one is based on real-space renormalization and applies to the symmetry group $S_4$ \cite{miller_resource_2015}. Two further formalisms \cite{raussendorf_symmetry-protected_2017, raussendorf_measurement-based_2023} apply to Abelian symmetry groups, such as the group $\mathbb{Z}_2 \times \mathbb{Z}_2$ in question \cite{else_symmetry-protected_2012, Son_2012}. The earlier of these formalisms \cite{raussendorf_symmetry-protected_2017} is based directly on the MPS framework provided in \cite{else_symmetry-protected_2012}, which imported the group-cohomological characterization of SPT phases from condensed matter physics. We subsequently denote it as the MPS-formalism. The later formalism \cite{raussendorf_measurement-based_2023} is not based on MPS but instead has a coding theoretic flavour. We denote it as the CT-formalism. 

While the MPS-formalism and the CT-formalism have the same scope, at the present stage of development, the CT-formalism is better at making quantitative predictions.  The MPS-formalism, however, is better at building an intuition, and we therefore use it in this background section.

The main purpose of this section is to explain the notion of logical decoherence in SPT-MBQC---how it arises and how it is mitigated.

\paragraph{MBQC in correlation space.} To prepare, we briefly discuss MBQC in the matrix-product state (MPS) picture, a.k.a. MBQC in correlation space \cite{gross_novel_2007}. Together with the group cohomological description introduced to MBQC in \cite{else_symmetry-protected_2012}, it provides the basis of the present discussion.

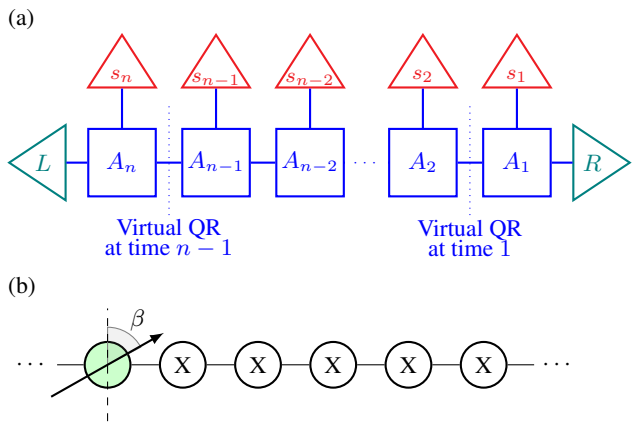
\begin{figure}
\centering
\begin{tabular}{l}
(a) \\ \begin{tikzpicture}[square/.style={regular polygon,regular polygon sides=4}, every text node part/.style={align=center}]
    \draw[teal, thick] (0.75, 0) -- (1.5, 0.5) -- (1.5, -0.5) -- cycle;
    \node[teal] at (1.2, 0) {$L$};
    \node[thick, blue, square, draw] (n) at (2.25, 0) {\phantom{$A_n$}};
    \node[blue] (n') at (2.25, 0) {$A_n$};
    \node[thick, blue, square, draw] (n-1) at (3.5, 0) {\phantom{$A_n$}};
    \node[blue] (n-1') at (3.5, 0) {$A_{n-1}$};
    \node[thick, blue, square, draw] (n-2) at (4.75, 0) {\phantom{$A_n$}};
    \node[blue] (n-2') at (4.75, 0) {$A_{n-2}$};
    \node[blue] (mid) at (5.5, 0) {\tiny $\cdots$};
    \node[thick, blue, square, draw] (2) at (6.25, 0) {\phantom{$A_n$}};
    \node[blue] (2') at (6.25, 0) {$A_{2}$};
    \node[thick, blue, square, draw] (1) at (7.5, 0) {\phantom{$A_n$}};
    \node[blue] (1') at (7.5, 0) {$A_{1}$};
    \draw[teal, thick] (8.25, 0.5) -- (9, 0) -- (8.25, -0.5) -- cycle;
    \node[teal] at (8.5, 0) {$R$};
    \draw[blue, thick] (1.51, 0) -- (n) -- (n-1) -- (n-2) -- (mid) -- (2) -- (1) -- (8.24, 0);
    \draw[dotted, blue] (2.875, 0.75) -- (2.875, -0.75);
    \node[blue] at (2.875, -1) {Virtual QR \\ at time $n-1$};
    \draw[dotted, blue] (6.875, 0.75) -- (6.875, -0.75);
    \node[blue] at (6.875, -1) {Virtual QR \\ at time $1$};

    \node[Red] at (2.25, 1.15) {\footnotesize$s_n$};
    \draw[Red, thick] (1.8, 1) -- (2.7, 1) -- (2.25, 1.7) -- cycle;

    \node[Red] at (3.5, 1.1) {\footnotesize$s_{n-1}$};
    \draw[Red, thick] (3.05, 1) -- (3.5+0.45, 1) -- (3.5, 1.7) -- cycle;

    \node[Red] at (4.75, 1.1) {\footnotesize$s_{n-2}$};
    \draw[Red, thick] (4.75-0.45, 1) -- (4.75+0.45, 1) -- (4.75, 1.7) -- cycle;

    \node[Red] at (6.25, 1.15) {\footnotesize$s_{2}$};
    \draw[Red, thick] (6.25-0.45, 1) -- (6.25+0.45, 1) -- (6.25, 1.7) -- cycle;

    \node[Red] at (7.5, 1.15) {\footnotesize$s_{1}$};
    \draw[Red, thick] (7.5-0.45, 1) -- (7.5+0.45, 1) -- (7.5, 1.7) -- cycle;
    
    \draw[blue, thick] (2.25, 0.46) -- (2.25, 0.985);
    \draw[blue, thick] (3.5, 0.46) -- (3.5, 0.985);
    \draw[blue, thick] (4.75, 0.46) -- (4.75, 0.985);
    \draw[blue, thick] (6.25, 0.46) -- (6.25, 0.985);
    \draw[blue, thick] (7.5, 0.46) -- (7.5, 0.985);
\end{tikzpicture}

\\ (b) \\ 
\begin{tikzpicture}
        \node[] (-1a) at (-6, 0) {$\cdots$};
        \node[circle, draw, thick, fill=green!20!] (0a) at (-5, 0) {\phantom{X}};
        \node[circle, draw, thick] (1a) at (-4, 0) {X};
        \node[circle, draw, thick] (2a) at (-3, 0) {X};
        \node[circle, draw, thick] (3a) at (-2, 0) {X};
        \node[circle, draw, thick] (4a) at (-1, 0) {X};
        \node[circle, draw, thick] (5a) at (0, 0) {X};
        \draw [shift={(-5, 0)}, gray, fill, fill opacity=0.1] (0,0) -- (28:0.5) arc (28:90.:0.5) -- cycle;
        \draw[-latex, thick] (-5.75, -0.425) -- (-4.25, 0.425);
        \draw[dashed] (-5, -0.75) -- (-5, 0.75);
        \node[] (0) at (-4.6, 0.6) {$\beta$};
        \node[] at (-6, 0) {\phantom{$\cdots$}};
        \node[] (6a) at (1, 0) {$\cdots$};
        \draw[] (-1a) -- (0a) -- (1a) -- (2a) -- (3a) -- (4a) -- (5a) -- (6a);
\end{tikzpicture}
\end{tabular}
\caption{\label{MPS}MBQC resource state and measurement pattern. (a) MPS representation of MBQC. (b) Implementation of a logical operation, consisting of a symmetry-breaking measurement and trailing oblivious wire.}
\end{figure}

All information about a given MBQC is contained in the overlaps $\langle \textbf{s}|\Phi\rangle$ between the post-measurement local state $|\textbf{s}\rangle = \bigotimes_i |s_i,\beta_i\rangle$, with $\textbf{s}=(s_1,..,s_N)$ the measurement record and the angles $\beta_i$ specifying the measurement bases at the sites $i$, and the MBQC resource state $|\Phi\rangle$. The MPS representation for this overlap is shown in Fig.~\ref{MPS} (a). In this picture, the MBQC-simulated quantum register is located on the virtual (=horizontal) links of the MPS network. The components $\langle s_i,\beta_i|A_i =: A_i[s_i]$ of the MPS tensor $A_i$ represent the quantum gates that successively act on the simulated quantum register.

For suitable resource states $|\Phi\rangle$, if the measurement basis is right, then all tensor components $ A_i[s_i]$ are unitary; and furthermore, those unitaries are the same up to an outcome-dependent Pauli operator $\Sigma(s_i)$,
\begin{equation}\label{idiTens}
A_i[s_i] = \Sigma(s_i)U_i(\beta_i).
\end{equation}
If this applies, then $U_i$ is the unitary simulated by the measurement of spin (or block) $i$. The random outcome-dependent Pauli operator $\Sigma(s_i)$, the so-called byproduct operator, is taken care of by forward-propagation and corresponding adaptation of subsequent measurement bases \cite{raussendorf_one-way_2001}.

The 1D cluster state is an example for the above formalism. Namely, whenever the measurement basis is an eigenbasis of an operator $\cos \beta X + \sin \beta Y$, for some angle $\beta$, the logical operation affected by local measurement is
$A_i[s_i]_\text{cluster} = X^{s_i}\, H\exp\left(-i\frac{\beta}{2} Z\right)$, $\forall i$.

\paragraph{Symmetry protected quantum wire.} The stepping stone to understanding logical processing in computational phases of quantum matter is the phenomenon of symmetry-protected quantum wire \cite{else_symmetry-protected_2012}. The result is that quantum wire, i.e., the ability to shuttle quantum information from one end of a spin chain to the other via local measurements, is uniform across suitable SPT phases. Any ground state in such a phase provides quantum wire, with perfect accuracy.

At the technical level, \cite{else_symmetry-protected_2012} shows that for special symmetry protected topological phases in 1D, namely those resulting from an Abelian symmetry group and a corresponding projective representation that is maximally non-commuting, any ground state $|\psi\rangle$ in the phase has an MPS representation with tensors $A$ of the form
\begin{equation}\label{Afact}
A^{(\psi)}[s_i] = C[s_i] \otimes B^{(\psi)}_\text{junk}[s_i].
\end{equation}
Therein, the tensor components are taken in the symmetry-respecting basis $s$ (i.e., the basis that commutes with the symmetry action). The matrices $C[s_i]$, acting on the so called `logical subspace' of the virtual space, are unitary and elements of a finite group, and they are constant across the phase. The so-called junk matrices $B^{(\alpha)}_\text{junk}[s_i]$, acting on the complementary junk space, are completely unspecified, and account for the difference between ground states in the phase.

Given the tensor product decomposition in correlation space between the logical and junk subspaces, we can simply use the logical subspace to store and teleport quantum information. As long as measurements are made in the symmetric basis, the evolution in Eq.~\eqref{Afact}.

Yet, the type of symmetry protected processing available is limited -- only the discrete group $\langle C[s_i]\rangle$ of gates can be enacted, in a random but heralded fashion. This amounts to quantum wire, i.e., deterministically executing the identity operation, with the matrices $C[s_i]$ playing the role of correctable byproduct operators.\smallskip

\paragraph{Logical decoherence.} It is at this stage that we encounter the main difficulty: to have more general processing than wire, the measurement basis has to be tilted away from the symmetry-respecting basis. But then, the tensor components $A[s_i]$ change, and the tensor product structure of Eq.~\eqref{Afact} disappears. As a consequence, measurement then entangles the logical with junk register, effectively leading to decoherence in the former. This is the phenomenon of logical decoherence.

\paragraph{Gate operation.} Logical decoherence in computational phases of quantum matter can be eliminated to arbitrary precision in a resource-efficient manner. Here we summarize the method of \cite{raussendorf_symmetry-protected_2017}, as it applies to the cluster phase of interest. The basic construction for implementing a logical rotation is the measurement pattern shown in Fig.~\ref{MPS} (b).
Therein, the trailing wire piece is called `oblivious wire'. It is similar to the symmetry-protected wire of \cite{else_symmetry-protected_2012}, but we consider the averaged effect over all measurement records, with the correction due to the byproduct operators taken into account. 

The effect of this averaging is that the junk subsystem is driven towards a fixed point $\rho_\text{fix}$, while the logical subsystem remains unaffected. In short, oblivious wire creates reproducible conditions of the computationally favorable factorized form. Its effect in combination with the symmetry-breaking measurement is as follows. We assume that the logical register is initially in a factorized state $\tau_\text{log} \otimes \rho_\text{fix}$. Then, the measurement at angle $\beta$ from the symmetry-respecting basis implements a logical gate, but at the cost of entangling the  logical and the junk subsystem. Finally, the trailing oblivious wire restores the factorization to $\tau'_\text{log} \otimes \rho_\text{fix}$. The effective logical action ${\cal{V}}$, as a function of the measurement angle $\beta$, is  \cite{raussendorf_measurement-based_2023}:
\begin{equation}\label{exaV}
{\cal{V}}_\beta = \frac{1+\sigma}{2}  \left[ \exp\left(-i \frac{\beta}{2} T \right)\right] +  \frac{1-\sigma}{2}  \left[ \exp\left(i \frac{\beta}{2} T \right)\right].
\end{equation}
Therein, the brackets $[\cdot]$ denote superoperators, and $T=X$ or $Z$, depending on the site to which the symmetry-breaking measurement is applied. The channel ${\cal{V}}_\beta$ acts on the logical subspace of the virtual system---the same space the byproduct operators $C[s_i]$ in Eq.~\eqref{Afact} act. 

This is, for measurement angles $\beta\neq 0$, a nontrivial logical action. However, it is also, for all $|\sigma|<1$, a non-unitary action leading to logical decoherence. To make this explicit, we may rewrite ${\cal{V}}_\beta$ as a unitary $U_{\text{log}}$ followed by a decoherent channel ${\cal{D}}(\epsilon):=(1-\epsilon)[1]+\epsilon[T]$, adding phase or spin-flip noise, respectively.
\begin{equation}\label{Dephase}
{\cal{V}}_\beta = {\cal{D}}\left(\frac{\epsilon}{4}\right) \left[ \exp\left(-i \frac{\beta_\text{log}}{2} T \right)\right].
\end{equation}
At this point, the prefactor of $1/4$ in ${\cal{D}}(\epsilon/4)$ appears arbitrary. It is inserted to maintain later consistency with the Frobenius norm as our general measure of error (see Appendix~\ref{app:norm}).

In Eq.~\eqref{Dephase}, the logical rotation angle $\beta_\text{log}$ is given by 
\begin{equation}\label{LogPar1}
\tan \beta_\text{log} = \sigma \tan \beta,
\end{equation}
and the error parameter $\epsilon$ is
\begin{equation}
\label{LogPar2}
    \epsilon = 2 \left(1-\sqrt{1-(1-\sigma^2)\sin^2\beta}\right). 
\end{equation}

The significance of the relations (\ref{LogPar1}), (\ref{LogPar2}) is that the logical rotation angle $\beta_\text{log}$ is, to leading order, linear in $\beta$ whereas the error rate $\epsilon$ is only quadratic in $\beta$,
\begin{equation}\label{SmallBeta}
\begin{array}{rcl}
\beta_\text{log} &=& \displaystyle{\sigma \beta + O(\beta^3),}\\
\epsilon &=& \displaystyle{\left(1-\sigma^2\right) \beta^2 + O(\beta^4).}
\end{array}
\end{equation}
The conclusion of Eq.~\eqref{SmallBeta} is that {\em{for small non-zero measurement angles $\beta$, gate action wins over decoherence.}} This is the basic property upon which uniformity of computational power across computational phases of matter rests. It is exploited in the splitting technique discussed below.

Before getting there, we discuss a further consequence of Eq.~\eqref{LogPar1}. It describes the logical response, specified by the rotation angle $\beta_\text{log}$ of a logical unitary gate, to the  symmetry-breaking local measurement at angle $\beta$. Ref.~\cite{raussendorf_symmetry-protected_2017} defines a computational order parameter $\nu$ characterizing this response,
\begin{equation}\label{nuDef}
\nu:=\lim_{\beta \rightarrow 0}\frac{\beta_\text{log}}{\beta},
\end{equation}
and expresses it in terms of the junk matrices of Eq.~\eqref{Afact}. With Eq.~\eqref{LogPar1}, we find that the physical order parameter $\sigma$ and the computational order parameter $\nu$ agree,
\begin{equation}\label{nu}
\sigma = \nu.
\end{equation}
This prediction is tested in Experiment \#2.
\smallskip

{\em{Splitting}} \cite{raussendorf_symmetry-protected_2017} is a fundamental technique for overcoming logical decoherence in computational phases of quantum matter. From Eq.~\eqref{SmallBeta} it follows that, if the measurement angle $\beta$ is cut in half, the logical rotation angle $\beta_\text{log}$ is cut in half too, but the deviation from unitarity is reduced by a factor of four. Thus, if an operation ${\cal{V}}[\beta]$ is replaced by two operations ${\cal{V}}[\beta/2]$, sufficiently spaced apart to make them independent, the combined deviation from unitarity is reduced by a factor of two, while the total rotation angle remains unchanged. 

Analogously, splitting a CPTP map ${\cal{V}}[\beta]$ into $m$ successive maps ${\cal{V}}[\beta/m]$ reduces the overall error by a factor of $m$. The deviation from unitarity can thus be made arbitrary small, at the expense of computational resources. This feature is crucial for extending MBQC from special resource states into SPT phases surrounding them.

The tradeoff between accuracy and resource consumption is described by a scaling relation for the logical error $\epsilon_m$ with $m$ \cite{raussendorf_symmetry-protected_2017, adhikary_counter-intuitive_2023},
\begin{equation}\label{FundScal}
\epsilon_m = \frac{1}{m}\kappa\, \beta^2 + O(1/m^2).
\end{equation}

A brief clarification regarding the error metric is warranted. For technical reasons~\cite{adhikary_counter-intuitive_2023}, the logical error is defined—up to an overall scalar factor—as the Frobenius norm of the difference between the target logical unitary and the implemented logical channel (See Appendix \ref{app:norm} for details).

The $\kappa$ in Eq.~\eqref{FundScal} measures the quality of the resource state. The lower the value of $\kappa$ the better the resource state. For an optimal state, such as the cluster state, $\kappa=0$, i.e., no logical error is incurred whatever the rotation angle.

For all symmetric resource states, $\kappa$ depends only on an order parameter (string order in the 1D $\mathbb{Z}_2\times \mathbb{Z}_2$ cluster phase) and its two-point correlation functions \cite{Adhikary_2021, adhikary_counter-intuitive_2023}. When the algorithmically non-trivial (i.e., symmetry-breaking) measurements are sufficiently far apart, then $\kappa$ only depends on the order parameter itself \cite{raussendorf_measurement-based_2023},
\begin{equation}\label{kasi}
\kappa=\frac{1-\sigma^2}{\sigma^2}.
\end{equation}
The consequence of the scaling relation Eq.~\eqref{FundScal} is that, as long as the value of $\kappa$ is finite, computation can proceed. The larger $\kappa$, the larger the splitting number $m$ has to be to compensate, hence the more costly the computation. Close to the boundary of a phase, computations therefore become very costly, but still all the same computations can be performed as in the fixed point of the phase (the cluster state in the $\mathbb{Z}_2 \times \mathbb{Z}_2$ case).\medskip

\smallskip

\emph{The correlated regime.} Upon closer inspection, the SPT-MBQC resource states are, in addition to the string order parameter $\sigma$, characterized by a correlation length $\xi$. It is set by the detrimental residual entanglement, and measures how quickly the string order parameters anchored at two distinct sites  of the 1D lattice de-correlate. When the spacing $\Delta$ between symmetry-breaking measurements comparable to or  less that $\xi$, then individual such measurements no longer give rise to individual logical operations. Rather, multiple measurements create a logical effect jointly. This is the correlated regime. It has to date been avoided, except in \cite{adhikary_counter-intuitive_2023}, due to the additional complications in treating it. 

The scaling relation Eq.~\eqref{FundScal} also holds in the correlated regime, with a value of $\kappa$ that is modified by the two-point correlation function
\begin{equation}\label{eq:sigma2l}
\sigma^2(l):=\langle Z_kX_{k+1}I_{k+2}..X_{k+l-1}Z_{k+l} \rangle,\;\;l\; \text{even}.
\end{equation}
of the string order parameter. In the translation-invariant bulk, it only depends on the difference $l$ between starting and end positions, not on the absolute position $k$. It is found that~\cite{adhikary_counter-intuitive_2023}
\begin{equation}
\kappa = \frac{1}{\sigma^2} \left( 1 -\sigma^2 + 2 \sum_{j=1}^{m-1}\left(\sigma^2(j\Delta) -\sigma^2\right)\right).
\end{equation}
When $\Delta\gg \xi$, it follows that $\sigma^2(j\Delta) \longrightarrow \sigma^2$ for all $j\geq 1$, and we recover Eq.~(\ref{kasi}).\smallskip

Two opposing effects compete in the correlated  regime. Putting two symmetry-breaking measurements closer together  increases the error of the combined logical operation. On the other hand, tighter spacing means more splitting in the same space, hence reduced error. 
As it turns out, under fairly general conditions the second factor wins \cite{adhikary_counter-intuitive_2023}. The correlated regime is the computationally most efficient. 

For our experiments, we use appropriate linear subsets of IBM's superconducting 127-qubit \texttt{ibm\_quebec} device. Errors are mitigated via measurement-error mitigation for the readout \cite{nation_scalable_2021}  as well as Pauli twirling \cite{wallman_noise_2016} and dynamic decoupling \cite{ezzell_dynamical_2023} for the 2-qubit echoed-cross resonance gates. We restrict ourselves to using unbiased error mitigation techniques, and as such do not use techniques such as zero-noise extrapolation \cite{giurgica-tiron_digital_2020}. The quantum circuits were written and simulated using the Qiskit SDK \cite{javadi-abhari_quantum_2024}. Error parameters of the devices are given in Appendix \ref{app:errors}.

\section{Experiment \#0: Resource State Preparation}\label{sec:exp0}

\subsubsection{Purpose}
For experimental demonstration of the phenomenology of computational phases of quantum matter, we require a family of resource states with tunable computational order. To this end, we consider a local transformation of the cluster state as a variational ansatz for the ground states of Eq.~\eqref{eq:interpolationH}:
\begin{equation}\label{eq:variationalansatz}
  \begin{array}{rcl}
  \ket{\Psi(\theta)} &=& \displaystyle{\bigotimes_{i=2}^{n-1}M_i(\theta)\ket{C_{n}}} \vspace{2mm}\\ 
  &\coloneqq & \displaystyle{\left(\bigotimes_{i=2}^{n-1} \cos(\frac{\theta}{2})I_i + \sin(\frac{\theta}{2})X_i\right)\ket{C_{n}}}
  \end{array}
\end{equation}
This can also be understood as the state obtained from applying imaginary time evolution $\exp(\beta  \sum_{i=2}^{n-1}X_i)$ with $\tanh\beta = \tan\frac{\theta}{2}$ to the cluster state.

Notably, this ansatz reproduces the ground-state structure predicted by first-order perturbation theory applied to the interpolation Hamiltonian of Eq.~\eqref{eq:interpolationH} (See Appendix \ref{app:perttheory}). It preserves the $\mathbb{Z}_2 \times \mathbb{Z}_2$ symmetry that defines the cluster phase under consideration and remains short-range entangled, ensuring its suitability for the theoretical framework developed in~\cite{raussendorf_measurement-based_2023, adhikary_counter-intuitive_2023}.

In addition, we note that this ansatz predicts the existence of a phase transition, though in the incorrect location at $\alpha = \arctan(2)$. From the experimental side, it is also easy to prepare as no entangling operations are required beyond the controlled-$Z$ gates used to prepare $\ket{C_n}$.

To approximate the ground state for a given $\alpha$, we look for the parameter $\theta$ which minimizes the energy:
\begin{equation}\label{eq:Hexpect}
    \bra{\Psi(\theta)}H(\alpha)\ket{\Psi(\theta)} = -\cos(\alpha)\sum_{i=1}^n \langle K_i \rangle_\theta - \sin(\alpha)\sum_{i=2}^{n-1}\langle X_i \rangle_\theta
\end{equation}

\subsubsection{Setup}
The expectation values of Eq.~\eqref{eq:Hexpect} can be evaluated by preparing the $n$-qubit cluster state, then applying the following probabilistic ancilla gadget that performs the nonunitary transformation on each site:

\begin{equation}\label{eq:M_igadget}
    M_i(\theta) = \begin{quantikz}[align equals at=1.5]
        \lstick{} & \targ & \qw & \qw
        \\ \lstick{$\lvert + \rangle$} & \ctrl{-1} & \gate{R_y(\theta)} & \meter{Z=+1}
    \end{quantikz}
\end{equation}

We can then measure $X_i, K_i$ on the resulting state. This scheme is inefficient as it is contingent on the exponentially small probability of obtaining the correct measurement outcomes on the $n-2$ ancilla qubits to obtain the full $n$-qubit state. However, through liberal use of symmetries of the cluster state and half-teleportation identities, we can equivalently obtain the expectation values from deterministic, small circuits. For example, the measurement of the magnetic field reduces to a single-qubit circuit:
\begin{equation}\label{eq:Xmeascircuit}
    \begin{quantikz}
        \lstick{$\ket{+}$} & \gate{R_y(\theta)} & \meter{Z}
    \end{quantikz}
\end{equation}
from which we can extract $\langle X_i\rangle_\theta = \langle Z \rangle$.

The boundary $i = 1, 2, n-1, n$ cluster stabilizers can be calculated from:
\begin{equation}\label{eq:Kboundmeascircuit}
    \begin{quantikz}
          \lstick{$\ket{0}$} & \gate{R_y(\theta)} & \meter{Z}
    \end{quantikz}
\end{equation}
with $\langle K_i\rangle_\theta = \langle Z\rangle$ and the bulk cluster stabilizers can be calculated from:
\begin{equation}\label{eq:Kbulkmeascircuit}
    \begin{quantikz}
          \lstick{$\ket{+}$} & \gate{R_y(\theta)} & \meter{Z_i}
    \end{quantikz}
\end{equation}
\begin{equation}\label{eq:Kbulkmeascircuit2}
    \begin{quantikz}
        \lstick{$\ket{+}$} & \ctrl{1} & \gate{Z^{s_0}} & \gate{R_y(\theta)} & \meter{Z_{i-1}}
          \\ \lstick{$\ket{+}$} & \control{} & \gate{Z^{s_i}} & \gate{H} & \gate{R_y(\theta)} & \meter{Z_{i+1}}
    \end{quantikz}
\end{equation}
with $s_0$ determined via coin flip and $\langle K_i\rangle_\theta = \langle (-1)^{s_{i-1} + s_{i} + s_{i+1}}\rangle$. The derivations for these compiled VQE circuits is given in Appendix \ref{app:vqecircuits}.

The correctness of the above circuits can be established through comparison with the expectation values derived from explicit calculation:
\begin{equation}\label{eq:Xi}
    \langle X_i \rangle_\theta = \sin(\theta)
\end{equation}
\begin{equation}\label{eq:Ki}
    \langle K_i \rangle_\theta = \begin{cases}
        \cos(\theta) & i = 0, 1, n-1, n
        \\ \cos^2(\theta) & i = 2, 3, \ldots, n-2
    \end{cases}
\end{equation}

\subsubsection{Results}

\begin{figure}[htbp!]
    \centering
\begin{tabular}{l}
(a) \\
  \includegraphics[width=0.9\linewidth]{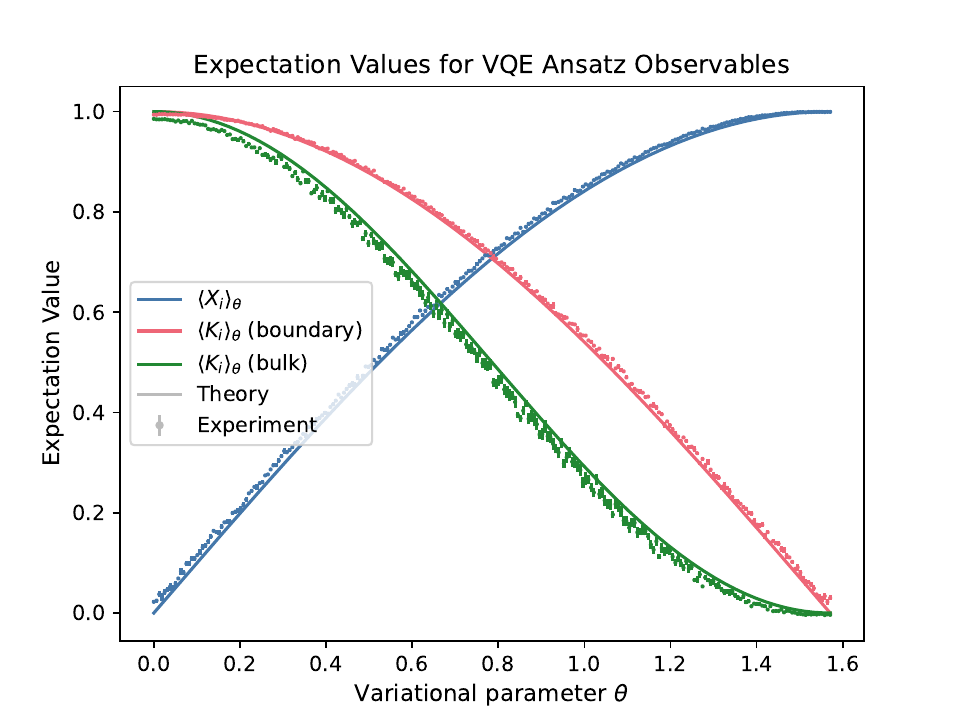} \\ 
(b) \\  
  \includegraphics[width=0.9\linewidth]{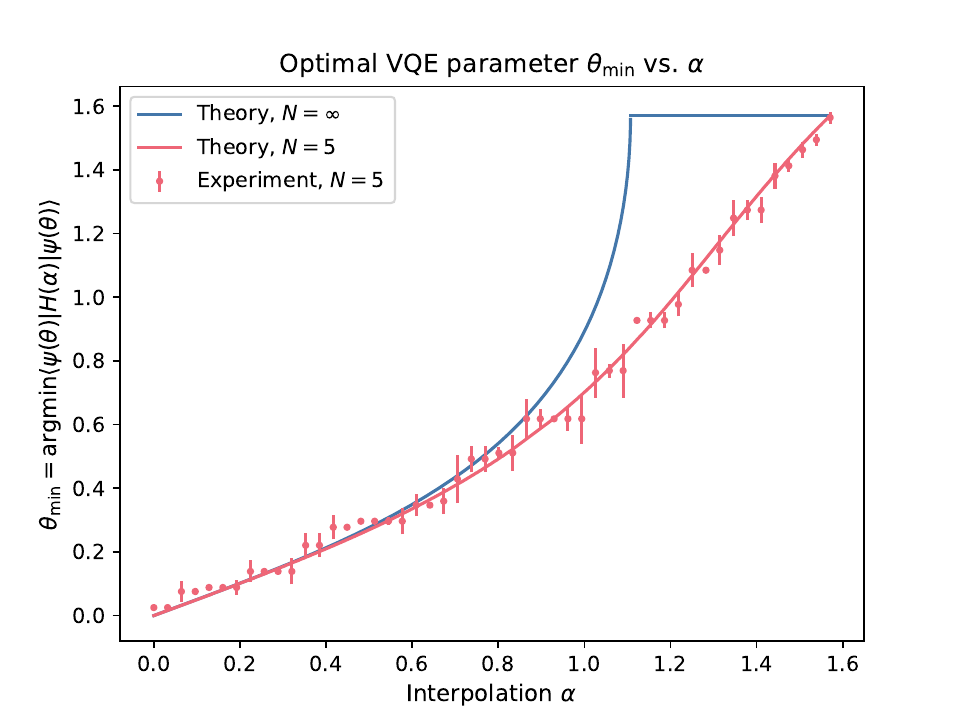}
\end{tabular}
    \caption{Experimental results for VQE (dots) with theory comparison (lines). (a) Expectation values of the energy terms. Individual points are obtained by measurements of the 1/2-qubit circuits of Eqs. \eqref{eq:Xmeascircuit} - \eqref{eq:Kbulkmeascircuit}, with 5000 shots per point. Error bars represent the standard error of the mean. (b) Optimization of the variational parameter depending on the Hamiltonian interpolation $\alpha$. We plot the experimental optimization for a 5-site chain, as well as theory results for the 5 and infinite-size chains (obtained by optimizing over the analytical results of Eqs. \eqref{eq:Xi}, \eqref{eq:Ki}). Error bars are estimated from the range of $\theta$ which fall within the minimum energy up to uncertainty. }
    \label{fig:VQE-results}
\end{figure}

We begin by performing the variational procedure for approximating the ground states of Eq.~\eqref{eq:interpolationH} via the ansatz state of Eq.~\eqref{eq:variationalansatz}. The measurements of the energy terms in the Hamiltonian as a function of the variational parameter $\theta$ and subsequent optimization are given in Fig.~\ref{fig:VQE-results}.

We see that generally the expectation values and optimal VQE parameters obtained experimentally agree with the calculated results. We also note the phase transition observed in the infinite size limit (see Appendix \ref{app:ansatzphasetransition} for the calculation), where for $\alpha \geq \arctan(2)$, $\theta_{\text{min}} = \frac{\pi}{2}$ and the variational ansatz state becomes the product state.

\section{Experiment \#1: Logical Decoherence}\label{sec:exp1}

\subsubsection{Purpose}

The first experiment tests the logical operations performed with resource states in the $\mathbb{Z}_2 \times \mathbb{Z}_2$ 1D cluster phase. Specifically, the dependence of the logical operations ${\cal{V}}_\beta$ of Eq.~\eqref{exaV} on the measurement angle $\beta$ is verified.

For this, we implement the measurement procedure on an imperfect yet symmetric resource state that would, on the perfect cluster state, be equivalent to the circuit of (i) preparation of the logical state $\ket{+}$, (ii) $z$-rotation about the angle $\beta$, (iii) measurement of $X$ or $Y$.

With Eq.~\eqref{exaV}, we obtain for the logical expectation values of $X$, $Y$,
\begin{equation}
\langle {X}\rangle = \cos \beta,\; \langle {Y}\rangle = \sigma\, \sin \beta. 
\end{equation}
Thus plotting the pairs $(\langle {X}\rangle (\beta), \langle {Y}\rangle (\beta))$ in the $X/Y$-plane with $\beta$ as a parameter, we obtain an ellipse with vertical half-axis of $\sigma$ and horizontal half-axis of 1. This is the prediction tested in the first experiment.

\subsubsection{Setup}

The technical hurdle at the onset is taking into account the non-unitary operators $M_i(\theta)$ of the deformed cluster state in the MBQC protocol. These are handled as follows. In the protocol, on each site $i$ we measure a local observable $O_i$ with eigenvalues $\pm 1$, with the eventual goal of evaluating $\langle O \rangle = \langle \bigotimes_{i=1}^n O_i \rangle$. Applying $M_i(\theta)$ on each site of the cluster state before this measurement is equivalent to locally measuring the operator:
\begin{equation}
    M_i^\dag(\theta) O_i M_i(\theta) = \lambda_i\dyad{\lambda_i}{\lambda_i} + \lambda_i^\perp \dyad{\lambda_i^\perp}{\lambda_i^\perp}
\end{equation}
which is still Hermitian. Thus, we may measure in the  eigenbasis $\set{\ket{\lambda_i}, \ket{\lambda_i^\perp}}$ which amounts to the basis transformation:
\begin{equation}
    U_i(\theta) = \dyad{0}{\lambda_i} + \dyad{1}{\lambda_i^\perp}
\end{equation}
We then post-process the measurement outcomes to obtain the expectation value on the deformed cluster state:
\begin{equation}\label{eq:deformedexpectation}
    \langle O \rangle = \sum_{j=1}^m \prod_{i=0}^n \lambda_i^j
\end{equation}
with the sum $j$ being over runs of the experiment. 

In the case of a wire basis measurement, we take $O_i = X_i$. Then, $M_i$ commutes (as it is $X$-type) and so $U_i(\theta)$ is just the identity - the only modification is the re-weighting of the measurement outcomes as per the eigenvalues in Eq.~\eqref{eq:deformedexpectation}. In the case of a rotated basis measurement, we take $O_i = \cos(\beta)X - \sin(\beta)Y$ and the unitary $U_i(\theta, \beta)$ becomes nontrivial.

\begin{figure}
    \begin{tabular}{l}
    (a) \\ 
        \begin{tikzpicture}
        \node[circle, draw, thick] (0a) at (-2, 0) {X};
        \node[circle, draw, thick] (1a) at (-1, 0) {X};
        \node[circle, draw, thick, fill=green!20!] (2a) at (0, 0) {\phantom{X}};
        \node[circle, draw, thick] (3a) at (1, 0) {X};
        \node[circle, draw, thick] (4a) at (2, 0) {\phantom{X}};
        \node[scale=0.85] at (2, 0) {X/Y};
        \draw [shift={(0, 0)}, gray, fill, fill opacity=0.1] (0,0) -- (28:0.5) arc (28:90.:0.5) -- cycle;
        \draw[-latex, thick, shift={(5, 0)}] (-5.75, -0.425) -- (-4.25, 0.425);
        \draw[dashed] (0, -0.75) -- (0, 0.75);
        \node[] (0) at (0.4, 0.6) {$\beta$};
        \draw[] (0a) -- (1a) -- (2a) -- (3a) -- (4a);
    \end{tikzpicture}
    \\ (b) \\ 
    \begin{quantikz}
        \lstick{$\ket{+}$} & \ctrl{1} & \qw & \qw  & \meter{X}
        \\ \lstick{$\ket{+}$} & \control{1} & \ctrl{1} & \qw & \meter{X}
        \\ \lstick{$\ket{+}$} & \ctrl{1} & \control{} & \gate{U(\theta, \beta)} & \meter{X}
        \\ \lstick{$\ket{+}$} & \control{1} & \ctrl{1} & \qw  & \meter{X}
        \\ \lstick{$\ket{+}$} & \qw & \control{} & \qw & \meter{X/Y}
    \end{quantikz}
    \end{tabular}
\caption{\label{eq:5qubitcircuit}(a) 5-qubit MBQC scheme with a single symmetry breaking measurement of angle $\beta$, used to verify the relation between logical gate action and decoherence. (b) 5-qubit quantum circuit with equivalent logical action, which is run on the IBM device. This circuit (and all others) was drawn using the Quantikz package \cite{kay_tutorialquantikzpackage_2023}.}
\end{figure}
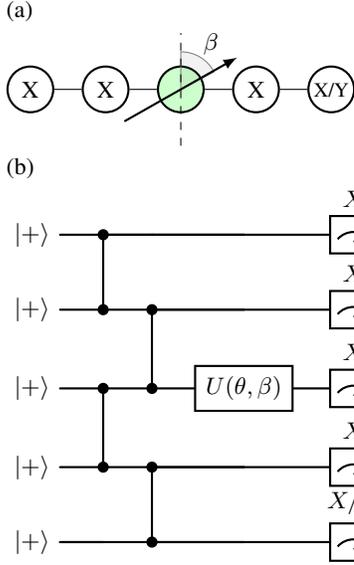

With this in mind, we consider a 5-qubit MBQC scheme, depicted in Fig.~\ref{eq:5qubitcircuit}(b).
This circuit corresponds to (i) the preparation of a cluster state with logical input $\ket{+}$. The deformation $M_i(\theta)$ is captured in the basis transformation on the third qubit - this is where we have the nontrivial rotated basis measurement inducing (ii) a $z$-rotation by angle $\beta$. Elsewhere, we measure in the wire/$X$-basis. Note that the native measurement basis of the experimental hardware is in the $Z$-basis, so this $X$-basis measurement is achieved by pre-measurement applications of the Hadamard gate.
(iii) On the last site, we measure $X_5$/$Y_5$ to perform state tomography on the logical state, wherein the expectation values $\langle X\rangle(\beta), \langle Y \rangle(\beta)$ can be computed from the measurement outcome (taking byproduct operators into account).

\subsubsection{Results}

For various values of the interpolation parameter $\alpha$, we find the corresponding minimizing VQE parameter $\theta$. We then run the circuit of \eqref{eq:5qubitcircuit} for various values of the logical rotation $\beta \in [0, \pi]$ to sweep out the predicted ellipses in the pairs $(\langle X\rangle (\beta), \langle Y\rangle (\beta))$.

\begin{figure}[htbp!]
    \centering
    \includegraphics[width=\linewidth]{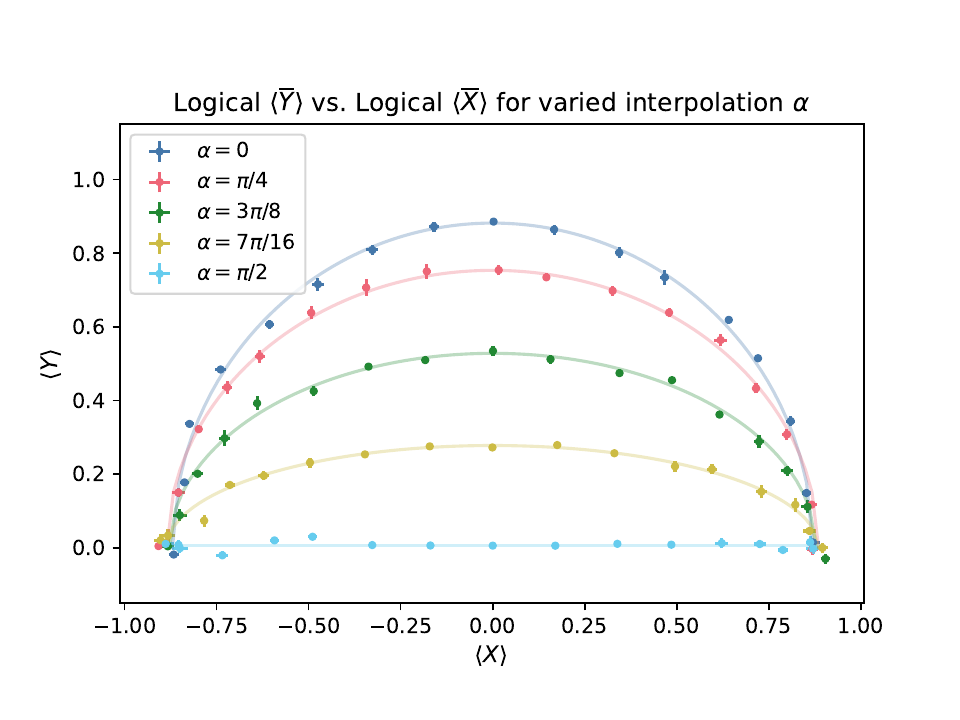}
    \caption{Experimental results for the logical expectation values $\langle X\rangle, \langle Y \rangle$, as a function of the Hamiltonian interpolation parameter $\alpha$ (which specifies the resource state) and the implemented rotation angle $\beta \in [0, \pi]$. Each point corresponds to 10000 shots for $\langle X \rangle, \langle Y \rangle$. Error bars represent the standard error of the mean. Solid lines represent best-fit ellipses to the data.}
    \label{fig:decoherence-results}
\end{figure}

Indeed, in Fig.~\ref{fig:decoherence-results} we observe ellipses with a uniform half axis of $1$, and a decaying vertical half axis as we increase the interpolation parameter $\alpha$. The vertical half axis corresponds to the string order parameter $\sigma$. As $\alpha$ increases from $\alpha = 0$ (the cluster state) to $\alpha = \pi/2$ (the product state), the string order parameter $\sigma$ decreases and we see a corresponding decay in the vertical half-axis from its maximal to zero value. Both the horizontal and vertical axes are reduced from their theoretical maximum value of unity due to the presence of device noise. In the following experiment, we confirm via independent measurement that the value of the half-axes indeed correspond to the string order parameter.
 
\section{Experiment \#2: String Order is Computational Order}\label{sec:exp2}

\subsubsection{Purpose}

Symmetry-protected order is 1D is string order \cite{den_nijs_preroughening_1989, tasaki_quantumliquid_1991, perez_stringorder_2008}, and comes with a string order parameter $\sigma$. It turns out that the string order parameter has direct computational significance, cf. Eqs.~\eqref{exaV} and \eqref{LogPar1}. Namely, the string order parameter determines how much the angle $\beta_\text{log}$ of the logical rotation evoked by a single local measurement is reduced compared to the measurement angle $\beta$. This relation is predicted in Eq.~(\ref{LogPar1}), 
$$
\tan \beta_\text{log} = \sigma \tan \beta,
$$
and we test this prediction here in Experiment \#2.

We recall from Eq.~(\ref{nuDef}) that a computational order parameter $\nu$ was defined in \cite{raussendorf_symmetry-protected_2017}, 
$\nu:={\displaystyle\lim_{\beta \to 0}} \frac{\beta_\text{log}}{\beta}$. If Eq.~(\ref{LogPar1}) is experimentally confirmed for the present setting, it also confirms that $\sigma=\nu$; i.e. that the physical and the computational order parameter agree.

\subsubsection{Setup}
The measurement of the computational order is obtained via the same methods as the previous experiment - the ratio of the logical expectation values $\langle Y\rangle(\beta)/\langle X \rangle(\beta) = \tan(\beta_{\text{log}})$ yields a line in $\tan(\beta)$, whose slope yields $\nu$. The independent measurement of the string order parameter $\sigma$ is done by measuring the expectation value of the appropriate Pauli string for the resource state directly. For our 5-qubit resource state with $\mathbb{Z}_2 \times \mathbb{Z}_2$-symmetry, the string order parameter is:
\begin{equation}
\sigma = \langle Z_3X_4Z_5\rangle.
\end{equation}
This is equivalent to the boundary cluster stabilizer measurement performed in Experiment \# 0, which is experimentally measured by the circuit given by \eqref{eq:Kboundmeascircuit}.

\subsubsection{Results}

\begin{figure}[htbp!]
    \centering
    \begin{tabular}{l}
    (a) \\
        \includegraphics[width=0.9\linewidth]{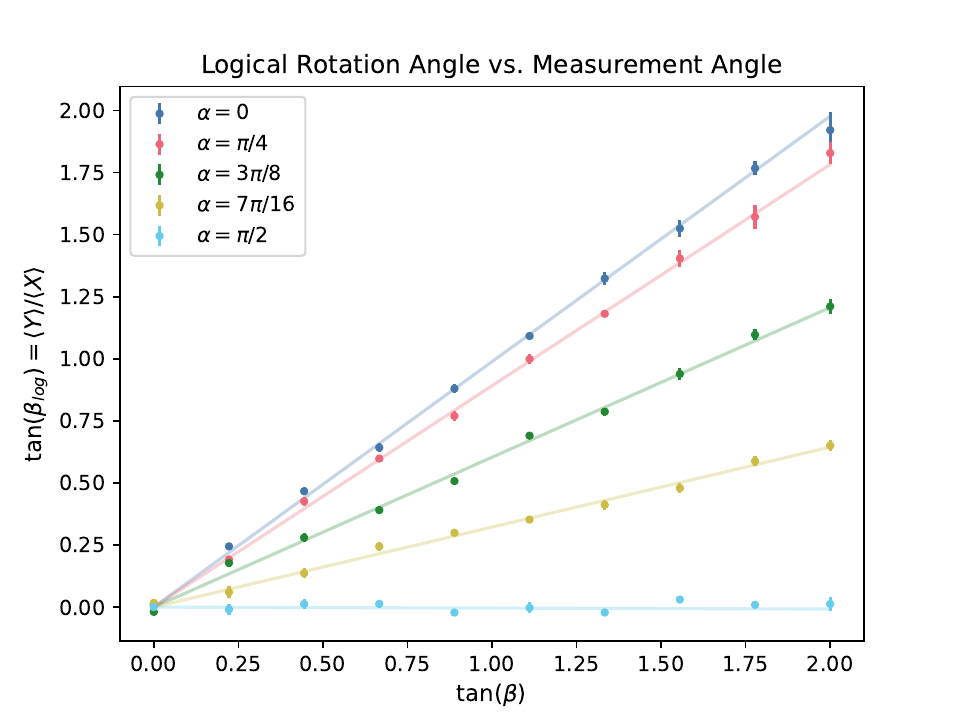}
        \\
        (b)
        \\
        \includegraphics[width=0.9\linewidth]{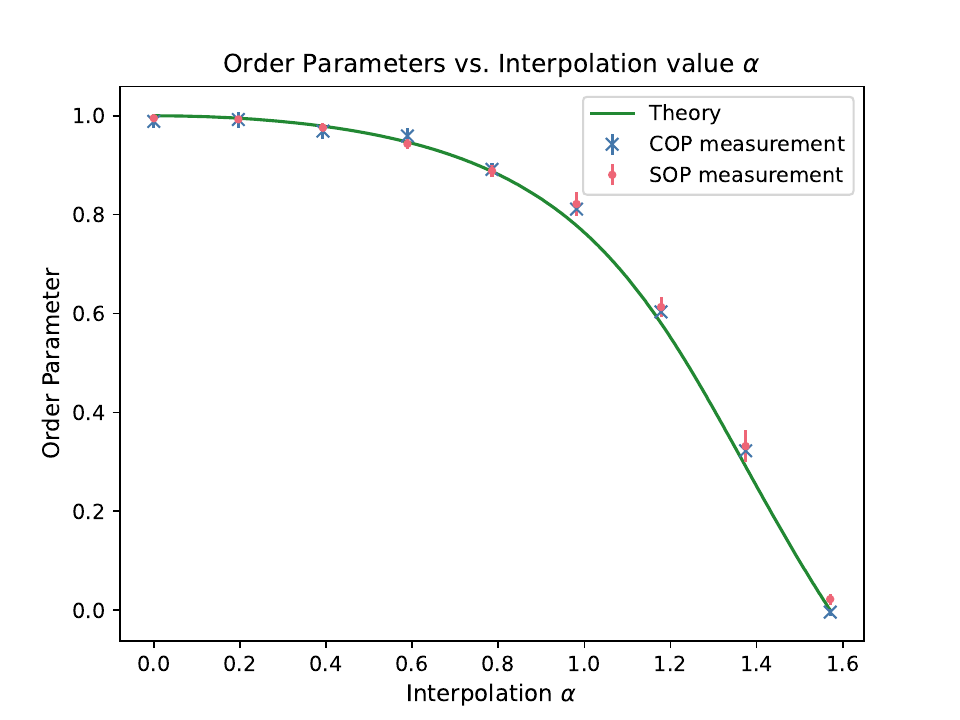}
    \end{tabular}
    \caption{Experimental results for demonstration of string order equals computational order. (a) Ratio of the logical $Y, X$ expectation values plotted against the tangent of the rotation angle $\beta$. Solid lines represent best-fit linest to the data. (b) The linear fits of (a) yield values of the computational order parameter (COP, blue crosses). The string order parameter (SOP, pink dots) is obtained from the appropriate cluster state stabilizer value measured in Experiment \#0. The theory curve (green) is obtained from the analytical expression in Eq.~(\ref{eq:Ki}). Error bars for the COPs are obtained from the variance in the optimal paramaeters of the least squares fit. Error bars for the SOP are obtained by combining the statistical error in the stabilizer measurement with the variance in $\langle K_i\rangle_\theta$ arising from determining the optimal $\theta_{\text{min}}$ for a given $\alpha$.}
    \label{fig:SOPCOP-results}
\end{figure}

The direct measurement of the string order parameter is provided by the boundary $\langle K_i\rangle_\theta$ curve in Fig.~\ref{fig:VQE-results} (for variationally determined $\theta)$. The computational order is obtained from the ratio of expectation values in Fig.~\ref{fig:decoherence-results}, which is depicted in Fig.~\ref{fig:SOPCOP-results}(a). The two are compared in Fig.~\ref{fig:SOPCOP-results}(b). We observe that up to statistical error, the two order parameters are measured to be equivalent for a wide range of the interpolation parameter $\alpha$, thus confirming the prediction.

\section{Experiment \#3: Splitting Mitigates Logical Decoherence}\label{sec:exp3}

\subsubsection{Purpose}

This is the central experiment of the four presented. Here, we test the universal scaling relation Eq.~\eqref{FundScal}, specifically the inverse scaling of the logical error with the splitting number $m$. Splitting is the basic technique for mitigating logical decoherence in computational phases of quantum matter. 

With Eq.~\eqref{FundScal},  through $m$-fold splitting the amount of logical decoherence is reduced to \cite{raussendorf_symmetry-protected_2017,Adhikary_2021}
\begin{equation}\label{Dm}
{\cal{D}}_m \propto \frac{\beta^2}{m}.
\end{equation}
Thus, in the limit of large $m$, logical decoherence may be suppressed arbitrarily. We test Eq.~\eqref{Dm} for small values of $m$.

\subsubsection{Setup}

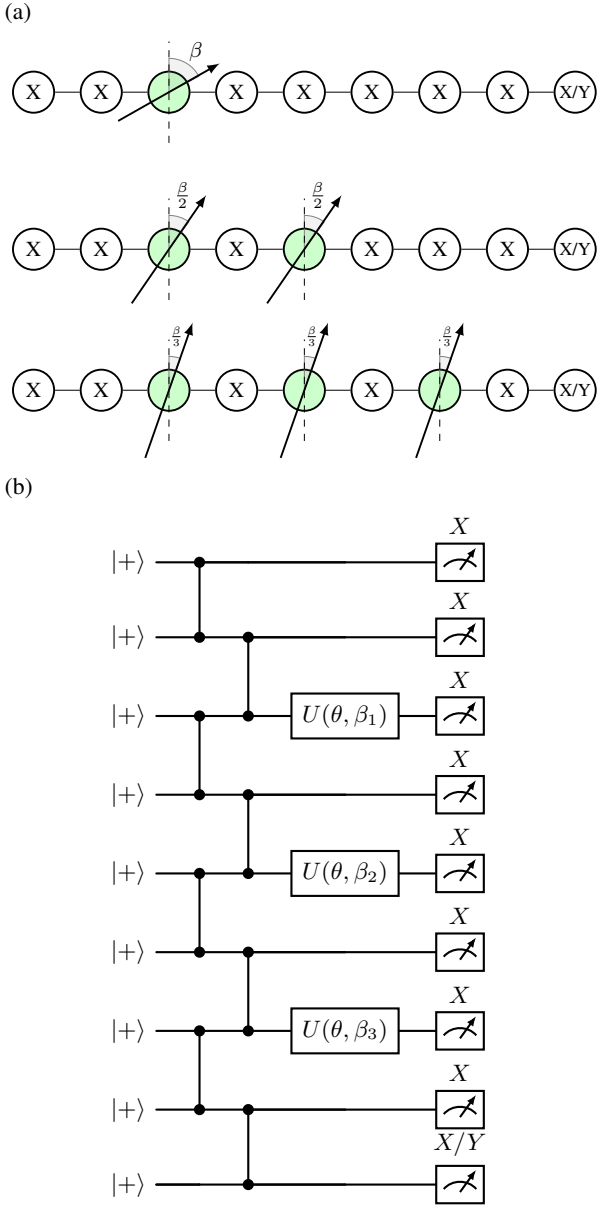
\begin{figure}
    \begin{center}
    \begin{tabular}{l}
    (a) \\ 
            \begin{tikzpicture}
        \node[scale=0.9] at (0, 4){
            \begin{tikzpicture}
                \node[circle, draw, thick] (2b) at (-3, 0) {X};
                \node[circle, draw, thick] (3b) at (-2, 0) {X};
                \node[circle, draw, thick, fill=green!20!] (4b) at (-1, 0) {\phantom{X}};
                \node[circle, draw, thick] (5b) at (0, 0) {X};
                \node[circle, draw, thick] (6b) at (1, 0) {X};
                \node[circle, draw, thick] (7b) at (2, 0) {X};
                \node[circle, draw, thick] (8b) at (3, 0) {X};
                \node[circle, draw, thick] (9b) at (4, 0) {X};
                \node[circle, draw, thick] (10b) at (5, 0) {\phantom{X}};
        
                \draw [shift={(-1, 0)}, gray, fill, fill opacity=0.1] (0,0) -- (28:0.5) arc (28:90.:0.5) -- cycle;
                \draw[-latex, thick] (-1.75, -0.425) -- (-0.25, 0.425);
                \draw[dashed] (-1, -0.75) -- (-1, 0.75);
                \node[] (0) at (-0.6, 0.6) {$\beta$};

                \node[scale=0.85] at (5, 0) {X/Y};
        
                \draw[] (2b) -- (3b) -- (4b) -- (5b) -- (6b) -- (7b) -- (8b) -- (9b) -- (10b);
            \end{tikzpicture}
        };
        \node[scale=0.9] at (0, 2){
            \begin{tikzpicture}
                \node[circle, draw, thick] (2b) at (-3, 0) {X};
                \node[circle, draw, thick] (3b) at (-2, 0) {X};
                \node[circle, draw, thick, fill=green!20!] (4b) at (-1, 0) {\phantom{X}};
                \node[circle, draw, thick] (5b) at (0, 0) {X};
                \node[circle, draw, thick, fill=green!20!] (6b) at (1, 0) {\phantom{X}};
                \node[circle, draw, thick] (7b) at (2, 0) {X};
                \node[circle, draw, thick] (8b) at (3, 0) {X};
                \node[circle, draw, thick] (9b) at (4, 0) {X};
                \node[circle, draw, thick] (10b) at (5, 0) {\phantom{X}};
        
                \draw [shift={(-1, 0)}, gray, fill, fill opacity=0.1] (0,0) -- (54:0.5) arc (54:90.:0.5) -- cycle;
                \draw[-latex, thick] (-1.55, -0.8) -- (-0.45, 0.8);
                \draw[dashed] (-1, -0.75) -- (-1, 0.75);
                \node[] (0) at (-0.8, 0.8) {$\frac{\beta}{2}$};

                \draw [shift={(1, 0)}, gray, fill, fill opacity=0.1] (0,0) -- (54:0.5) arc (54:90.:0.5) -- cycle;
                \draw[-latex, thick] (0.45, -0.8) -- (1.55, 0.8);
                
                \draw[dashed] (1, -0.75) -- (1, 0.75);
                \node[] (0) at (1.2, 0.8) {$\frac{\beta}{2}$};

                \node[scale=0.85] at (5, 0) {X/Y};
        
                \draw[] (2b) -- (3b) -- (4b) -- (5b) -- (6b) -- (7b) -- (8b) -- (9b) -- (10b);
            \end{tikzpicture}
        };
        \node[scale=0.9]at (0, 0){
            \begin{tikzpicture}
                \node[circle, draw, thick] (2b) at (-3, 0) {X};
                \node[circle, draw, thick] (3b) at (-2, 0) {X};
                \node[circle, draw, thick, fill=green!20!] (4b) at (-1, 0) {\phantom{X}};
                \node[circle, draw, thick] (5b) at (0, 0) {X};
                \node[circle, draw, thick, fill=green!20!] (6b) at (1, 0) {\phantom{X}};
                \node[circle, draw, thick] (7b) at (2, 0) {X};
                \node[circle, draw, thick, fill=green!20!] (8b) at (3, 0) {\phantom{X}};
                \node[circle, draw, thick] (9b) at (4, 0) {X};
                \node[circle, draw, thick] (10b) at (5, 0) {\phantom{X}};
        
                \draw [shift={(-1, 0)}, gray, fill, fill opacity=0.1] (0,0) -- (70:0.5) arc (70:90.:0.5) -- cycle;
                \draw[-latex, thick] (-1.35, -1) -- (-0.65, 1);
                \draw[dashed] (-1, -0.75) -- (-1, 0.75);
                \node[scale=0.75] (0) at (-0.89, 0.8) {$\frac{\beta}{3}$};
        
                \draw [shift={(1, 0)}, gray, fill, fill opacity=0.1] (0,0) -- (70:0.5) arc (70:90.:0.5) -- cycle;
                \draw[-latex, thick] (0.65, -1) -- (1.35, 1);
                \draw[dashed] (1, -0.75) -- (1, 0.75);
                \node[scale=0.75] (0) at (1.11, 0.8) {$\frac{\beta}{3}$};
        
                \draw [shift={(3, 0)}, gray, fill, fill opacity=0.1] (0,0) -- (70:0.5) arc (70:90.:0.5) -- cycle;
                \draw[-latex, thick] (2.65, -1) -- (3.35, 1);
                \draw[dashed] (3, -0.75) -- (3, 0.75);
                \node[scale=0.75] (0) at (3.11, 0.8) {$\frac{\beta}{3}$};

                \node[scale=0.85] at (5, 0) {X/Y};
        
                \draw[] (2b) -- (3b) -- (4b) -- (5b) -- (6b) -- (7b) -- (8b) -- (9b) -- (10b);
            \end{tikzpicture}
        };
    \end{tikzpicture}
\\ (b) \\ 
    \begin{tikzpicture}
    \node[] at (-3.8, 0) {};
    \node[] at (0, 0) {
    \begin{quantikz}
        \lstick{$\ket{+}$} & \ctrl{1} & \qw & \qw & \meter{X}
        \\ \lstick{$\ket{+}$} & \control{} & \ctrl{1} & \qw & \meter{X}
        \\ \lstick{$\ket{+}$} & \ctrl{1} & \control{} & \gate{U(\theta, \beta_1)} & \meter{X}
        \\ \lstick{$\ket{+}$} & \control{} & \ctrl{1} & \qw& \meter{X}
        \\ \lstick{$\ket{+}$} & \ctrl{1} & \control{} & \gate{U(\theta, \beta_2)} & \meter{X}
        \\ \lstick{$\ket{+}$} & \control{} & \ctrl{1} & \qw & \meter{X}
        \\ \lstick{$\ket{+}$} & \ctrl{1} & \control{} & \gate{U(\theta, \beta_3)} & \meter{X}
        \\ \lstick{$\ket{+}$} & \control{} & \ctrl{1} & \qw & \meter{X}
        \\ \lstick{$\ket{+}$} & \qw & \control{} & \qw & \meter{X/Y}
    \end{quantikz}};
    \end{tikzpicture}
    \end{tabular}
    \end{center}
    \caption{\label{Circ1} (a) 9-qubit MBQC scheme with varied splittings of the logical rotation angle, used to verify the scaling of the logical decoherence. (b) 9-qubit quantum circuit with equivalent logical action (for appropriate choices of $(\beta_1, \beta_2, \beta_3)$ in the three cases), which is run on the IBM device.}
\end{figure}

We consider a nine-qubit MBQC scheme, implemented via the circuit displayed in Fig.~\ref{Circ1}.

The structure is identical to that of the first experiment, with the lengthening of the chain allowing for the splitting of the $z$-rotation across multiple sites. We consider the three cases of (i) $(\beta_1, \beta_2, \beta_3) = (\beta, 0, 0)$ (no splitting), (ii) $(\beta_1, \beta_2, \beta_3) = (\beta/2, \beta/2, 0)$ (splitting in half), (iii) $(\beta_1, \beta_2, \beta_3) = (\beta/3, \beta/3, \beta/3)$ (splitting into three). For each case, we can measure $\langle {X}\rangle(\beta), \langle {Y}\rangle(\beta)$ as in the first experiment, and from this extract the loss in purity of the logical state as:
\begin{equation}\label{eq:LP}
    \text{LP}_m(\beta) := 1 - \text{Tr}({\rho}^2) = \frac{1 - \left(\langle {X}\rangle(\beta)\right)^2 - \left( \langle {Y}\rangle(\beta)\right)^2}{2}.
\end{equation}
Note that $\langle Z \rangle$ can be neglected from this expression as $\langle Z \rangle = 0$ for the initial input state $\ket{+}$ and this is preserved by the logical evolution of $z$-rotations. We may then (for small angles $\beta$) measure the scaling of $\text{LP}_m$, where the curvature is predicted to scale (Eq.~\eqref{Dm}) as $\frac{1}{m}$ in the splitting $m$.

For the initial state used, the purity loss under a single-site $z$-rotation takes the form \(\frac{\epsilon}{2}(1-\frac{\epsilon}{4})\), which fixes it uniquely as a function of the error parameter of Eq.~\eqref{LogPar2}. In the small-\(\epsilon\) regime, reducing the purity loss and reducing \(\epsilon\) are equivalent. The benefit of purity loss over $\epsilon$ in the experimental setting is its simplicity of characterization - we need only measure the logical Pauli operators.

The circuit of Fig.~\ref{Circ1} lacks the usual MBQC primitive of adaptive measurements, as all measurements are performed at the end in a single step. Generically, this requires post-selection of the measurement outcomes, as ``incorrect'' intermediate measurement outcomes flip future rotation angles. However, in the case of uncorrelated measurements, the loss in purity is invariant under flips $\beta \leftrightarrow -\beta$ of rotation angles, as the error is quadratic in $\beta$ (Eq.~\eqref{SmallBeta}). Hence, post-selection can be avoided.

\subsubsection{Results}

\begin{figure}[htbp!]
\centering
\begin{tabular}{l}
(a) \\ 
    \includegraphics[width=0.9\linewidth]{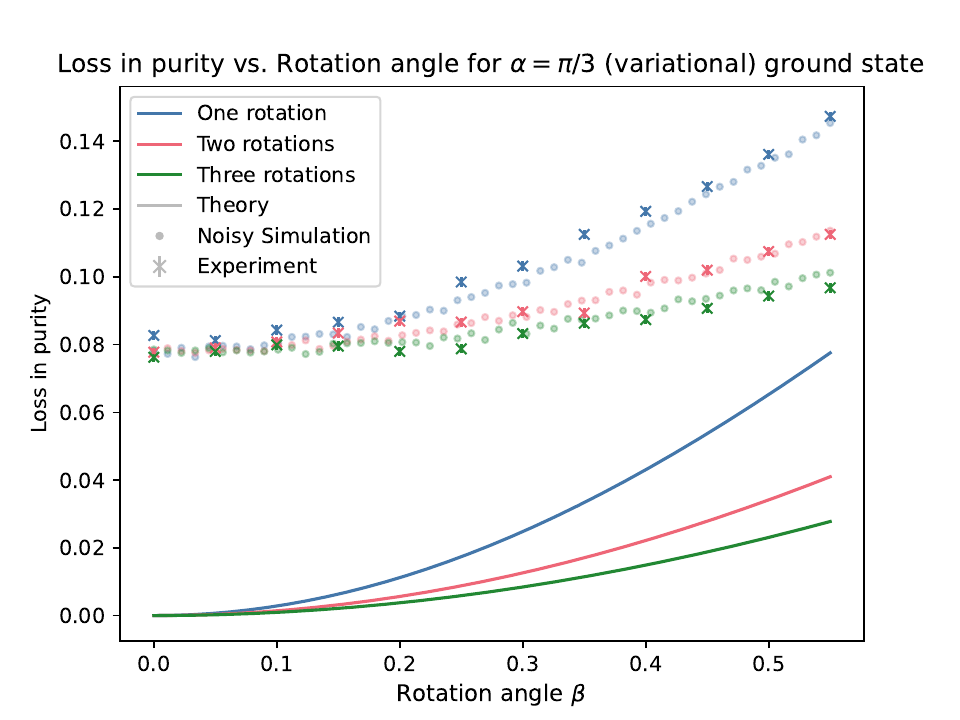} \\
    (b) \\
    \includegraphics[width=0.9\linewidth]{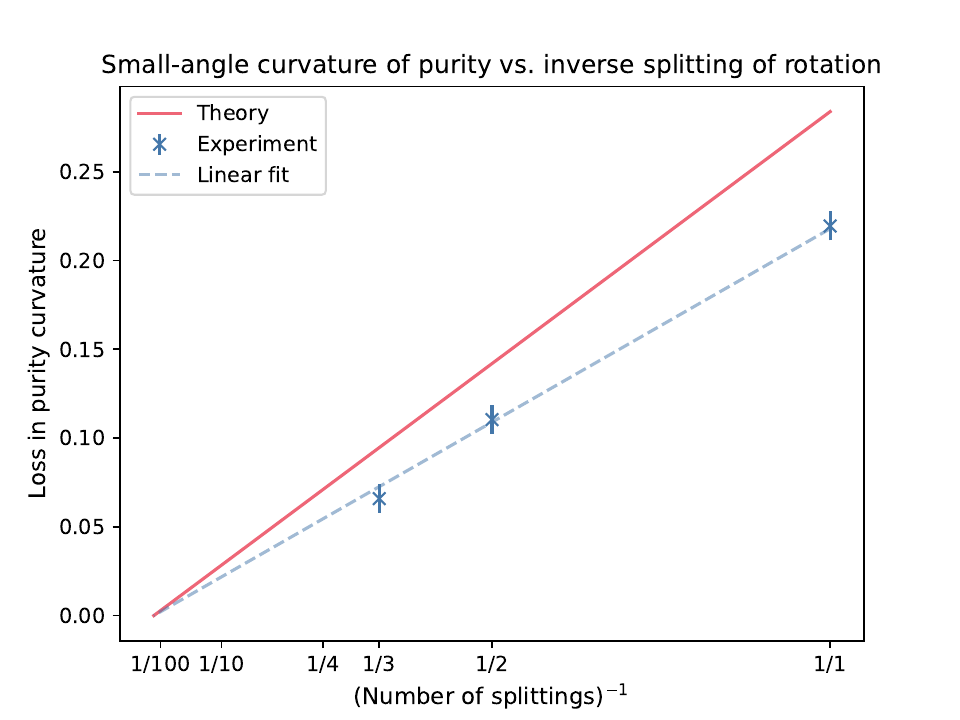}
\end{tabular}
    \caption{Experimental results for rotation splitting. (a) Taking the resource state to be the variational ansatz Eq.~\eqref{eq:variationalansatz} for interpolation parameter $\alpha = \pi/3$, we compare the loss in purity for small angles $\beta$ for (i) a single rotation (blue), (ii) two rotations (red), and (iii) three rotations (green). 240 runs of 10000 shots were taken and averaged per data point, with error bars corresponding to the standard error of the mean. Theory curves are calculated using the exact expressions for the expectation values, obtained by application of the channel of Eq.~\eqref{exaV}. The noisy simulation is obtained by simulating 4 million shots of the circuits with two-qubit gate depolarizing noise of strength $p = 0.0105$. (b) Curvatures of the small angle loss in purity of versus (inverse) number of splittings. The experimental points (blue crosses) are obtained by fitting the data in (a) by a quadratic with offset, and are fit by a best-fit line (dashed, blue), with slope $m = 0.218(7)$. The theory curve (solid, red) is a plot of the expressions in Eqs. \eqref{FundScal}, \eqref{kasi} with $\sigma$ analytically obtained from Eq.~\eqref{eq:Ki}.}
    \label{fig:rotation-split}
\end{figure}

Fig.~\ref{fig:rotation-split} (a) depicts the small-angle loss in purity for the three cases of rotation splitting. The loss in purity scales quadratically, though is offset from the theory prediction due to experimental noise, which is well-captured by a simple noise model involving a depolarizing channels applied uniformly to all two-qubit gates. In (b), we test the $\frac{1}{m}$ scaling of the decoherence by plotting the curvature of the loss in purity against the number of splittings. We find that the experimental data indeed obeys this scaling, as the curvatures form a line when plotted against the inverse number of splittings. The slope of this line is decreased compared to theory due to a uniform reduction of the curvatures due to device noise, but nontheless the prediction of the form of the scaling is verified.

\section{Experiment \#4: The Counterintuitive Regime}\label{sec:exp4}
\subsubsection{Purpose}

This is a refined experiment on splitting, in the correlated regime. This regime is entered when the separation between two symmetry breaking measurements becomes very small. Under these circumstances, multiple consecutive measurements give rise to a logical operation jointly. 

In spite of this complication, the fundamental scaling relation Eq.~\eqref{FundScal} persists. Only the value of $\kappa$ changes. A question that now arises is whether the correlated regime should be avoided by spacing the computationally relevant measurements (the symmetry-breaking ones) sufficiently far apart.

In Ref.~\cite{adhikary_counter-intuitive_2023} it has been shown that this is not the case. To the contrary, the {\em{densest}} packing, probing the correlated regime the deepest, is the most efficient!

Specifically, Theorem~2 in \cite{adhikary_counter-intuitive_2023} states that densest packing is the most efficient if (i) the 2-point correlator of the string order parameter is a convex function of distance, and (ii) the system is in the long chain limit.

Under these conditions, for the 1D $\mathbb{Z}_2 \times \mathbb{Z}_2$ cluster phase, the optimal spacing $\Delta_\text{opt}$ of same-type logical operations thus is 
\begin{equation}\label{opt}
\Delta_\text{opt} = 2.
\end{equation}
This prediction is tested in Experiment 4. 

However, on a finite-size quantum computer we cannot satisfy condition (ii) of theorem---the thermodynamic limit. Condition~(i) alone does not suffice to guarantee optimality of densest packing. For small system sizes, there are parameter regions in which the order of efficiency is inverted; see Fig.~\ref{fig:11qubitCIanalysis} in Appendix~\ref{app:CIregimevalidity}.

For the present experiment, we choose a family of resource states that satisfy the convexity condition (i) and for which simulation predicts the optimality of densest packing. This prediction we compare to experiment.

\subsubsection{Setup}
Although the local deformation of the cluster state given in Eq.~\eqref{eq:variationalansatz} is sufficient for probing the effects of logical decoherence and mitigation thereof via splitting, it lacks a length scale. It is thus insufficient for demonstrating efficient methods for computation in the correlated regime. Therein, we introduce the $XX$-rotated cluster state:
\begin{equation}\label{eq:correlatedresource}
    \ket{\Omega(\phi)} = \prod_{i \in B}RX_{i}(\phi)\prod_{i=1}^{n-2}RXX_{i, i+2}(\phi)\ket{C_n}
\end{equation}
where $RXX_{i, j}(\phi) = e^{-i\frac{\phi}{2}X_iX_j}$ is a two-site rotation, $B = \set{3, n-2}$ are sites on the boundary of the bulk rotation region, and $RX_i(\phi) = e^{-i\frac{\phi}{2}X_i}$ are single site $X$-rotations (added to compensate for boundary effects). This state is still $\mathbb{Z}_2 \times \mathbb{Z}_2$ symmetric, but the presence of the next-nearest neighbour gates equips it with a length scale, allowing us to test predictions in the correlated regime. 

An assumption that must hold in making predictions concerning the correlated regime is the convex decay of the string order parameters/two-point correlators $\sigma^2(l)$ of Eq.~\eqref{eq:sigma2l} as a function of the distance $l$. For the state of Eq.~\eqref{eq:correlatedresource} on $n \geq 11$ spins, we indeed confirm this to be the case analytically, with:
\begin{equation}\label{ex:twopoint}
    \sigma^2(l) = \begin{cases}
        \cos^2(\phi) & l = 2
        \\ \cos^4(\phi) & l = 4, 6, \ldots
    \end{cases}.
\end{equation}
There is an additional subtlety in the argument of \cite{adhikary_counter-intuitive_2023} pertaining to finite-size effects, which we discuss in Appendix \ref{app:CIregimevalidity}. However, we nonetheless expect that for this (correlated) state we are able to demonstrate that the densest possible packing of logical rotations is the most optimal.

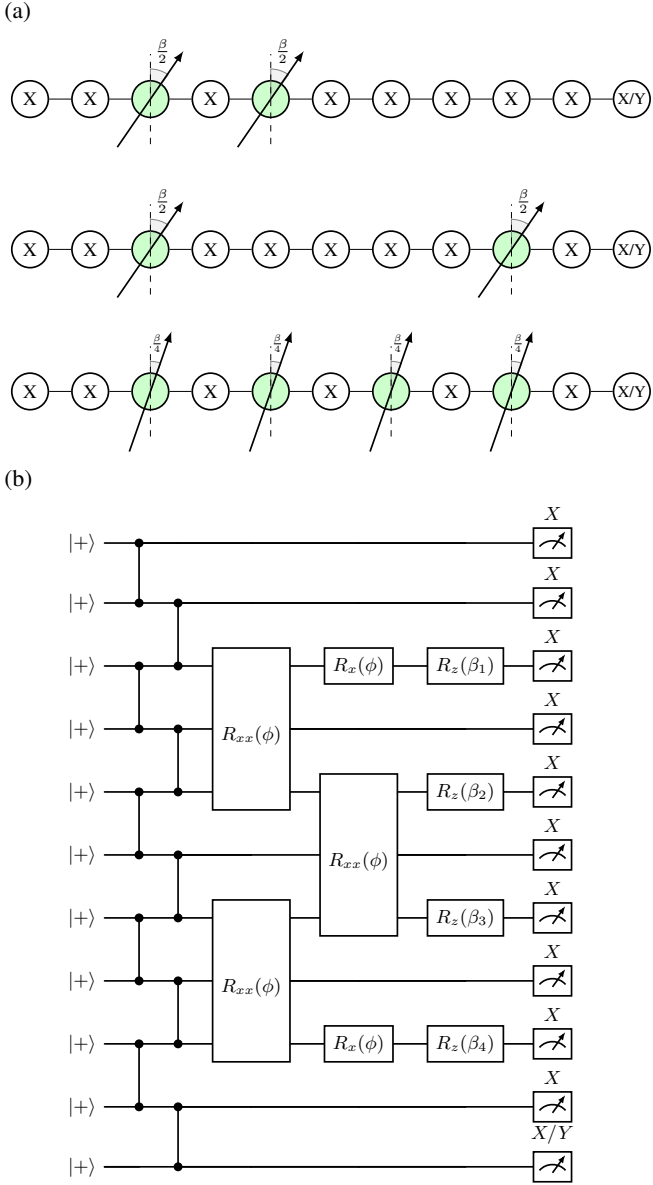
\begin{figure}
\begin{center}
\begin{tabular}{l}
(a) \\ 
        \begin{tikzpicture}
        \node[scale=0.8] at (0, 4){
            \begin{tikzpicture}
                \node[circle, draw, thick] (2b) at (-3, 0) {X};
                \node[circle, draw, thick] (3b) at (-2, 0) {X};
                \node[circle, draw, thick, fill=green!20!] (4b) at (-1, 0) {\phantom{X}};
                \node[circle, draw, thick] (5b) at (0, 0) {X};
                \node[circle, draw, thick, fill=green!20!] (6b) at (1, 0) {\phantom{X}};
                \node[circle, draw, thick] (7b) at (2, 0) {X};
                \node[circle, draw, thick] (8b) at (3, 0) {X};
                \node[circle, draw, thick] (9b) at (4, 0) {X};
                \node[circle, draw, thick] (10b) at (5, 0) {X};
                \node[circle, draw, thick] (11b) at (6, 0) {X};
                \node[circle, draw, thick] (12b) at (7, 0) {\phantom{X}};
        
                \draw [shift={(-1, 0)}, gray, fill, fill opacity=0.1] (0,0) -- (54:0.5) arc (54:90.:0.5) -- cycle;
                \draw[-latex, thick] (-1.55, -0.8) -- (-0.45, 0.8);
                \draw[dashed] (-1, -0.75) -- (-1, 0.75);
                \node[] (0) at (-0.8, 0.8) {$\frac{\beta}{2}$};

                \draw [shift={(1, 0)}, gray, fill, fill opacity=0.1] (0,0) -- (54:0.5) arc (54:90.:0.5) -- cycle;
                \draw[-latex, thick] (0.45, -0.8) -- (1.55, 0.8);
                
                \draw[dashed] (1, -0.75) -- (1, 0.75);
                \node[] (0) at (1.2, 0.8) {$\frac{\beta}{2}$};
                \node[scale=0.85] at (7, 0) {X/Y};
        
                \draw[] (2b) -- (3b) -- (4b) -- (5b) -- (6b) -- (7b) -- (8b) -- (9b) -- (10b) -- (11b) -- (12b);
            \end{tikzpicture}
        };
        \node[scale=0.8] at (0, 2){
            \begin{tikzpicture}
                \node[circle, draw, thick] (2b) at (-3, 0) {X};
                \node[circle, draw, thick] (3b) at (-2, 0) {X};
                \node[circle, draw, thick, fill=green!20!] (4b) at (-1, 0) {\phantom{X}};
                \node[circle, draw, thick] (5b) at (0, 0) {X};
                \node[circle, draw, thick] (6b) at (1, 0) {X};
                \node[circle, draw, thick] (7b) at (2, 0) {X};
                \node[circle, draw, thick] (8b) at (3, 0) {X};
                \node[circle, draw, thick] (9b) at (4, 0) {X};
                \node[circle, draw, thick, fill=green!20!] (10b) at (5, 0) {\phantom{X}};
                \node[circle, draw, thick] (11b) at (6, 0) {X};
                \node[circle, draw, thick] (12b) at (7, 0) {\phantom{X}};
        
                \draw [shift={(-1, 0)}, gray, fill, fill opacity=0.1] (0,0) -- (54:0.5) arc (54:90.:0.5) -- cycle;
                \draw[-latex, thick] (-1.55, -0.8) -- (-0.45, 0.8);
                \draw[dashed] (-1, -0.75) -- (-1, 0.75);
                \node[] (0) at (-0.8, 0.8) {$\frac{\beta}{2}$};

                \draw [shift={(5, 0)}, gray, fill, fill opacity=0.1] (0,0) -- (54:0.5) arc (54:90.:0.5) -- cycle;
                \draw[-latex, thick] (4.45, -0.8) -- (5.55, 0.8);
                
                \draw[dashed] (5, -0.75) -- (5, 0.75);
                \node[] (0) at (5.2, 0.8) {$\frac{\beta}{2}$};
                \node[scale=0.85] at (7, 0) {X/Y};
        
                \draw[] (2b) -- (3b) -- (4b) -- (5b) -- (6b) -- (7b) -- (8b) -- (9b) -- (10b) -- (11b) -- (12b);
            \end{tikzpicture}
        };
        \node[scale=0.8]at (0, 0){
            \begin{tikzpicture}
                \node[circle, draw, thick] (2b) at (-3, 0) {X};
                \node[circle, draw, thick] (3b) at (-2, 0) {X};
                \node[circle, draw, thick, fill=green!20!] (4b) at (-1, 0) {\phantom{X}};
                \node[circle, draw, thick] (5b) at (0, 0) {X};
                \node[circle, draw, thick, fill=green!20!] (6b) at (1, 0) {\phantom{X}};
                \node[circle, draw, thick] (7b) at (2, 0) {X};
                \node[circle, draw, thick, fill=green!20!] (8b) at (3, 0) {\phantom{X}};
                \node[circle, draw, thick] (9b) at (4, 0) {X};
                \node[circle, draw, thick, fill=green!20!] (10b) at (5, 0) {\phantom{X}};
                \node[circle, draw, thick] (11b) at (6, 0) {X};
                \node[circle, draw, thick] (12b) at (7, 0) {\phantom{X}};
        
                \draw [shift={(-1, 0)}, gray, fill, fill opacity=0.1] (0,0) -- (70:0.5) arc (70:90.:0.5) -- cycle;
                \draw[-latex, thick] (-1.35, -1) -- (-0.65, 1);
                \draw[dashed] (-1, -0.75) -- (-1, 0.75);
                \node[scale=0.75] (0) at (-0.89, 0.8) {$\frac{\beta}{4}$};
        
                \draw [shift={(1, 0)}, gray, fill, fill opacity=0.1] (0,0) -- (70:0.5) arc (70:90.:0.5) -- cycle;
                \draw[-latex, thick] (0.65, -1) -- (1.35, 1);
                \draw[dashed] (1, -0.75) -- (1, 0.75);
                \node[scale=0.75] (0) at (1.11, 0.8) {$\frac{\beta}{4}$};
        
                \draw [shift={(3, 0)}, gray, fill, fill opacity=0.1] (0,0) -- (70:0.5) arc (70:90.:0.5) -- cycle;
                \draw[-latex, thick] (2.65, -1) -- (3.35, 1);
                \draw[dashed] (3, -0.75) -- (3, 0.75);
                \node[scale=0.75] (0) at (3.11, 0.8) {$\frac{\beta}{4}$};

                \draw [shift={(5, 0)}, gray, fill, fill opacity=0.1] (0,0) -- (70:0.5) arc (70:90.:0.5) -- cycle;
                \draw[-latex, thick] (4.65, -1) -- (5.35, 1);
                \draw[dashed] (5, -0.75) -- (5, 0.75);
                \node[scale=0.75] (0) at (5.11, 0.8) {$\frac{\beta}{4}$};
                \node[scale=0.85] at (7, 0) {X/Y};
        
                \draw[] (2b) -- (3b) -- (4b) -- (5b) -- (6b) -- (7b) -- (8b) -- (9b) -- (10b) -- (11b) -- (12b);
            \end{tikzpicture}
        };
        
        \end{tikzpicture}
\\ (b) \\
\begin{tikzpicture}
\node[] at (-4.1, 0) {};
\node[scale=0.8] at (0, 0){
    \begin{quantikz}
    \lstick{$\ket{+}$} & \ctrl{1} & \qw & \qw & \qw & \qw & \meter{X} \\
    \lstick{$\ket{+}$} & \control{} & \ctrl{1} & \qw & \qw & \qw & \meter{X} \\
    \lstick{$\ket{+}$} & \ctrl{1} & \control{} & \gate[3, label style = {yshift=-0.1cm}]{R_{xx}(\phi)} & \gate{R_x(\phi)} & \gate{R_z(\beta_1)} & \meter{X} \\
    \lstick{$\ket{+}$} & \control{} & \ctrl{1} & \linethrough & \qw & \qw & \meter{X} \\
    \lstick{$\ket{+}$} & \ctrl{1} & \control{} & & \gate[3, label style = {yshift=-0.1cm}]{R_{xx}(\phi)} & \gate{R_z(\beta_2)} & \meter{X} \\
    \lstick{$\ket{+}$} & \control{} & \ctrl{1} & \qw & \linethrough & \qw & \meter{X} \\
    \lstick{$\ket{+}$} & \ctrl{1} & \control{} & \gate[3, label style = {yshift=-0.1cm}]{R_{xx}(\phi)} &  & \gate{R_z(\beta_3)} & \meter{X} \\
    \lstick{$\ket{+}$} & \control{} & \ctrl{1} & \linethrough & \qw & \qw &  \meter{X} \\
    \lstick{$\ket{+}$} & \ctrl{1} & \control{} & & \gate{R_x(\phi)} & \gate{R_z(\beta_4)} & \meter{X} \\
    \lstick{$\ket{+}$} & \control{} & \ctrl{1} & \qw & \qw & \qw & \meter{X} \\
    \lstick{$\ket{+}$} & \qw & \control{} & \qw & \qw & \qw & \meter{X/Y}
\end{quantikz}
};
\end{tikzpicture}
\end{tabular}
\caption{\label{Circ2}(a) 11-qubit MBQC scheme with varied splitting and spacing to test the optimality of the counterintuitive regime/densest packing of symmetry-breaking measurements. (b) 11-qubit quantum circuit involving the creation + measurement of a resource state in the cluster phase (with longer-range correlations arising from then nearest neighbour rotation gates), which is run on the IBM device.}
\end{center}
\end{figure}

The circuit representing our 11-qubit MBQC scheme is shown in Fig.~\ref{Circ2}. 

We consider the three cases of (i) $(\beta_1, \beta_2, \beta_3, \beta_4) = (\beta/2, \beta/2, 0, 0)$ (two rotations with $\Delta = 2$), (ii) $(\beta_1, 0, 0, \beta_4) = (\beta/2, 0, 0, \beta/2)$ (two rotations with $\Delta = 6$), and (iii) $(\beta_1, \beta_2, \beta_3, \beta_4) = (\beta/4, \beta/4, \beta/4, \beta/4)$ (four rotations with $\Delta = 2$). For each, we measure the logical expectation values $\langle X\rangle(\beta), \langle Y\rangle(\beta)$, and compute the loss in purity of the logical state according to Eq.~\eqref{eq:LP}. The theory predicts that the loss in purity is ordered between the three cases as (i) $>$ (ii) $>$ (iii), with (ii) outperforming (i) as two uncorrelated ($\Delta = 6$) logical operations outperforms two correlated ($\Delta = 2$) operations, and (iii) outperforming both cases as it is the densest possible packing/splitting of rotations with $\Delta_{\text{opt}} = 2$.

The next-nearest neighbour $XX$ rotations appearing in the above circuit cannot be performed natively on the IBM hardware, so instead are compiled to nearest-neighbour gates using intermediate SWAP operations.

In contrast to Experiment \#3 where the loss in purity was invariant under flips of rotation angles, in the correlated regime probed here the sign of the rotations enters, and hence post-selection on the outcomes of the intermediate qubits 4/6/8 is required in lieu of adaptive measurements.

\subsubsection{Results}
\begin{figure}[htbp!]
    \centering
    \includegraphics[width=0.9\linewidth]{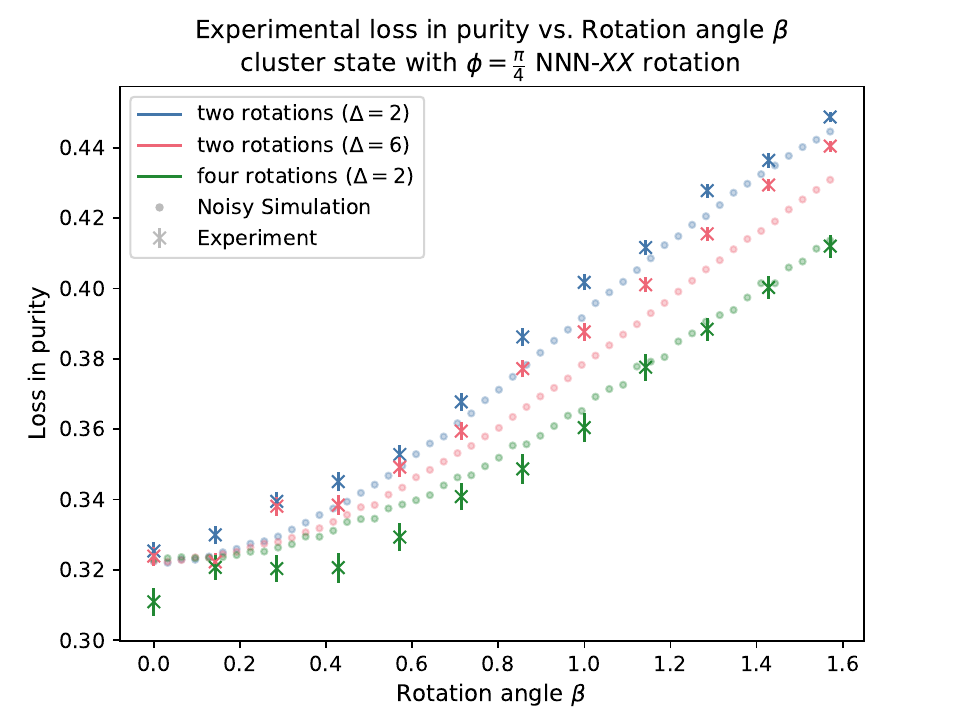}
    \caption{Experimental results for the counterintuitive regime. Using the $XX$-rotated cluster state with $\phi = \pi/4$ as the resource state, we compare (i) two $\Delta = 2$ rotations (blue), (ii) two $\Delta = 6$ rotations (red), and (iii) four $\Delta = 2$ rotations (green). 240 runs of 10000 shots each were taken and averaged per data point, with error bars corresponding to the standard error of the mean. The noisy simulation is obtained by simulating 4 million shots of the circuits with two-qubit gate depolarizing noise of strength $p = 0.0205$.}
    \label{fig:CI-results}
\end{figure}

We consider the $XX$-rotated cluster state of Eq.~\eqref{eq:correlatedresource} with $\phi = \pi/4$ as to maximize the decay of the string order parameters, and experimentally measure the loss in purity for the three cases for a range of total rotation angle $\beta$. The results are depicted in Fig.~\ref{fig:CI-results}. 

Overall, we see that across a range of rotation angles $\beta$ that the expected purity loss ranking of (i) $>$ (ii) $>$ (iii) is obeyed, though there is an inconsistency at $\beta = 0$ where the three datapoints are expected to coincide. The experimental noise again appears to be relatively well captured by the simple noise model of uniform depolarizing errors on all two-qubit gates. The baseline loss in purity is higher than in Experiment \#3 owing to the increased number of two-qubit gates (arising from the non-local $XX$ gates), and also possibly due to increased qubit relaxation errors due to longer circuit runtimes (that may arise in our noise model as a higher two-qubit gate error). Overall, the data gives highly suggestive experimental evidence for the counterintuitive regime of densest packing of rotations being the most efficient.

\section{Discussion}\label{sec:discussion}

In this paper, we have comprehensively tested theoretical predictions about computational phases of quantum matter. Our testbed was the one-dimensional $\mathbb{Z}_2 \times \mathbb{Z}_2$ cluster phase. We have performed four experiments, with the main one being the third--it verifies a scaling relation that guarantees uniform computational power across the phase. The first two experiments build towards the third, and the fourth experiment tests an implication of the scaling relation in the more complicated correlated regime, to which the scaling relation still applies.

In the first experiment, we tested whether the symmetric imperfections of the MBQC resource state impact the logical quantum operations in the way  predicted by theory. For measurement-induced logical rotation of a $|+\rangle$ state about the $z$-axis, the prediction is that expectation values $\langle X \rangle$, $\langle Y \rangle$ after rotation should lie on an ellipse, with horizontal half-axis 1 and vertical half-axis $\leq 1$. This quantifies the effect of logical decoherence on a single gate. Logical decoherence should be the largest for the Clifford gate $\exp(\mp i \frac{\pi}{4} Z)$, and vanish for Pauli gates $I$ and $Z$ (computational wire).  The measured data, displayed in Fig.~\ref{fig:decoherence-results}, confirms this behaviour.

The second experiment was concerned with the relation between physical and computational order, at the quantitative level. From the perspective of quantum computation, a computational order parameter is defined (cf.~Eq.~\eqref{nu}) by the efficiency with which a measurement angle translates into the corresponding rotation angle of the effected logical operation. From the perspective of physics, the $\mathbb{Z}_2\times \mathbb{Z}_2$ cluster phase is described by a non-vanishing string order parameter. Theory says that the computational order parameter and the physical string order parameter should be the same. Our experiment confirms this; see Fig.~\ref{fig:SOPCOP-results}.

The third and main experiment tested the universal scaling relation Eq.~\eqref{FundScal}, on which the uniformity of measurement-based computational power across symmetry protected phases rests. It says that the logical error of a quantum operation can be arbitrarily reduced by splitting this operation into many equal parts, each close to the identity. Specifically, $m$-fold splitting leads to $m$ fold reduction of logical error, at the expense of an $m$-fold increase of computational resources spent. This scaling relation is experimentally confirmed; see Fig.~\ref{fig:rotation-split}.

The fourth experiment tested an implication of the universal scaling relation in the correlated regime, namely that the densest packing of the algorithm-driving (hence symmetry breaking) measurements is the most efficient---in spite of the correlations. 
The corresponding data is presented in Fig.~\ref{fig:CI-results}, confirming, for the parameter values chosen, the prediction for the ordering of computational efficiencies. \smallskip

In sum, we have experimentally confirmed the uniformity of MBQC computational power across the one-dimensional cluster phase with $\mathbb{Z}_2 \times \mathbb{Z}_2$ symmetry. We have also confirmed theoretically predicted optimal computational strategies, which is the first such test. 

In the future, as quantum computational capacities grow, it is of interest to verify computational uniformity for computational phases with larger symmetry groups, both in 1D and 2D. Such phases have typically more---sometimes universal---computational power \cite{Stephen2024DualUnitary}, \cite{Daniel2020Archimedean,Stephen2019Subsystem, Raussendorf2019ComputationalPhase, Devakul2018FractalSPT}.

Another exciting prospect is the experimental mapping-out of computational phase diagrams---from known to unknown. A rich phase diagram in  intermediate range is the 2D topological $XZ$-star model of \cite{Herringer2025MBQC_SET}. It features several non-trivial computational phases, one of which is universal. 
\smallskip

\section*{Data Availability}

Experimental data and scripts used for analysis can be accessed at \cite{githubrepo_2025}.

\section*{Acknowledgements}
We thank Amrit Guha and Ruben Verresen for useful discussions. We thank the BC Quantum Algorithms Institute for providing access to IBMQ devices. RW acknowledges the financial support of the Natural Sciences and Engineering Research Council of Canada (NSERC) through a Canada Graduate Scholarships–Master’s (CGS-M) award. DB is funded by the Canada First Research Excellence Fund. AA and RR are funded by the Humboldt Foundation.

We acknowledge the use of IBM Quantum services for this work. The views expressed are those of the authors, and do not reflect the official policy or position of IBM or the IBM Quantum team.


\bibliography{main.bib}

\appendix

\onecolumngrid

\begin{appendices}


\setcounter{secnumdepth}{1}

\newpage
\section{Perturbation Theory for the VQE Ansatz}\label{app:perttheory}
\noindent
We consider the interpolating Hamiltonian
\begin{equation}
H(\alpha)
= -\cos(\alpha)\sum_{i=1}^{n} K_i
  - \sin(\alpha)\sum_{i=2}^{n-1} X_i ,
\label{eq:Halpha}
\end{equation}
where the operators \( K_i \) are the cluster-state stabilizer generators.

\medskip
\noindent For small angles, we expand
\begin{equation}
H(\alpha)
= H_0 + \alpha V + O(\alpha^2),
\label{eq:Hexpansion}
\end{equation}
with
\begin{align}
H_0 &= -\sum_{i=1}^{n} K_i,
\label{eq:H0} \\
V   &= -\sum_{i=2}^{n-1} X_i .
\label{eq:V}
\end{align}

\medskip
Let the unperturbed ground state be the \(n\)-qubit cluster state
\(|C_n\rangle\), satisfying
\begin{align}
K_i |C_n\rangle &= |C_n\rangle,
\label{eq:KiCn} \\
H_0 |C_n\rangle &= -n |C_n\rangle,
\label{eq:H0Cn} \\
E_0^{(0)} &= -n .
\label{eq:E0unpert}
\end{align}

Acting with \(X_i\) flips the eigenvalues of \(K_{i-1}\) and \(K_{i+1}\),
producing an excitation energy \(E_\Delta = 4\),
\begin{equation}
H_0 X_i |C_n\rangle = (-n + 4)\, X_i |C_n\rangle .
\label{eq:excited}
\end{equation}

\medskip
Using nondegenerate Rayleigh--Schrödinger perturbation theory,
the first-order corrections are
\begin{align}
E_0^{(1)} &= \langle C_n | V | C_n \rangle = 0,
\label{eq:E1} \\[6pt]
|0^{(1)}\rangle
&= \sum_{m \neq 0}
   \frac{\langle m | V | 0 \rangle}{E_0^{(0)} - E_m^{(0)}} |m\rangle
\nonumber \\
&= \frac{\alpha}{4} \sum_{i=2}^{n-1} X_i |C_n\rangle .
\label{eq:psi1}
\end{align}

\medskip
Hence, the (unnormalised) ground state to first order in \(\alpha\) is
\begin{align}
|\mathrm{GS}(\alpha)\rangle
&= \Bigg( 1
  + \frac{\alpha}{4}\sum_{i=2}^{n-1} X_i
   \Bigg)
  |C_n\rangle  + O(\alpha^2),
\label{eq:GState} \\[6pt]
E_0(\alpha)
&= -n + O(\alpha^2).
\label{eq:Ealpha}
\end{align}

\section{Derivations of VQE Circuits}\label{app:vqecircuits}
For the purposes of minimizing the ground state energy of the interpolating Hamiltonian, we wish to calculate the expectation values of $\langle K_i\rangle_\theta, \langle X_i\rangle_\theta$ for our resource state of Eq.~\eqref{eq:variationalansatz}:
\begin{equation}
  \begin{array}{rcl}
  \ket{\Psi(\theta)} &=& \displaystyle\bigotimes_{i=2}^{n-1}M_i(\theta)\ket{C_n} =  \displaystyle{\left(\bigotimes_{i=2}^{n-1} \cos(\frac{\theta}{2})I_i + \sin(\frac{\theta}{2})X_i\right)\ket{C_{n}}}
  \end{array}
\end{equation}

To construct this state, we start with the cluster state $\ket{C_n}$, and for each (non-boundary) qubit introduce an ancilla and apply the gadget of Eq.~\eqref{eq:M_igadget}:
\begin{equation}
    \begin{quantikz}[align equals at=1.5]
        \lstick{} & \targ & \qw & \qw
        \\ \lstick{$\lvert + \rangle$} & \ctrl{-1} & \gate{R_y(\theta)} & \meter{Z_{s_{z, i}}}
    \end{quantikz} =     \begin{cases}
 \cos(\frac{\theta}{2})I_i + \sin(\frac{\theta}{2})X_i = M_i(\theta) & Z_{s_{z, i}} = +1
        \\ X_i\left(\cos(\frac{\theta}{2})I_i - \sin(\frac{\theta}{2})X_i\right) = X_iM_i(-\theta) & Z_{s_{z, i}} = -1
    \end{cases}
\end{equation}
While the $X_i$ can be accounted for as a byproduct/error operator, the relative sign error in the case of a $-1$ outcome makes the above an exponentially inefficient implementation (as the correct state requires a $+1$ outcome on all $n-2$ ancilla qubits). To mitigte this probabilistic effect, we pull out a cluster state stabilizer with one end at the site with incorrect ancilla measurement outcome and the other end at the end of the chain (where the gadget is not applied). Then using:
\begin{equation}
    CX_{a, i}Z_i = Z_iZ_aCX_{a, i}
\end{equation}
we can use the $Z$ on the ancilla to flip the relative sign and implement the correct operation. Thus, locally, we consider the (now-deterministic) gadget:
\begin{equation}
    \begin{quantikz}[align equals at=1.5]
        \lstick{} & \gate{Z_{s_{z, i}}} & \targ & \qw & \qw
        \\ \lstick{$\lvert + \rangle$} & \qw & \ctrl{-1} & \gate{R_y(\theta)} & \meter{Z_{s_{z, i}}}
    \end{quantikz}
\end{equation}

With this starting point, we can start to compile the circuits to the efficient forms of Eqs. \eqref{eq:Xmeascircuit}-\eqref{eq:Kbulkmeascircuit2}. We discuss the $\langle X_i\rangle_\theta$ in detail, and provide the necessary additional identities involved for the similar $\langle K_i\rangle_\theta$ derivation.

For measuring $\langle X_i\rangle_\theta$, on each site we have:
\begin{equation}
\begin{quantikz}[align equals at=1.5]
    \lstick{} & \gate{Z_{s_{z, i}}} & \targ & \qw & \meter{X_{s_i}}
    \\ \lstick{$\lvert + \rangle$} & \qw & \ctrl{-1} & \gate{R_y(\theta)} & \meter{Z_{s_{z, i}}}
\end{quantikz}=
\begin{quantikz}[align equals at=1.5]
    \lstick{} & \meter{X_{s_i + s_{z, i}}}
    \\ \lstick{$\lvert + \rangle$} & \gate{Z^{s_i}} & \gate{R_y(\theta)} & \meter{Z_{s_{z, i}}}
\end{quantikz}=
\begin{quantikz}[align equals at=1.5]
    \lstick{} & \meter{X_{s_i + s_{z, i}}}
    \\ \lstick{$\lvert + \rangle$} & \gate{R_y(\theta)} & \meter{Z_{s_i + s_{z, i}}}
\end{quantikz}
\end{equation}
where in the first equality we commute the $X$-measurement with the controlled-$X$, and in the second equality we commute the $Z^{s_i}$ by virtue of the input ancilla state being real.

We define the shorthand:
\begin{equation}
    s_i' = s_{i} + s_{z, i}
\end{equation}
to be the sum of the physical and ancilla measurement outcome. 

We see that each ancilla has decoupled, and on the cluster state itself we are left with $X_{s_i + s_{z, i}}$-measurements on each site. This is simply a sequence of half-teleportations, which reduces to:
\begin{equation}
    \begin{quantikz}[align equals at=1]
        \ket{+} & \gate{Z^{s_1}} & \gate{H} & \gate{Z^{s_2'}} & \gate{H} & \cdots & \gate{H} & \gate{Z^{s_{n-1}'}} & \gate{H} & \gate{H} & \meter{X_{s_n}}
    \end{quantikz}=
    \begin{quantikz}[align equals at=1]
        \ket{+} & \qw & \meter{X_{s_0 + s_{2}' + \ldots + s_{n-2}' + s_n}}
    \end{quantikz}
\end{equation}

But this is simply the statement that $s_0 + s_{2}' + \ldots + s_{n-2}' + s_n = 0$, which follows automatically from the $\mathbb{Z}_2 \times \mathbb{Z}_2$ symmetry of the state. Thus, the circuit is extraneous and need not be run. To measure the local magnetic field, we only need to run the single-qubit disconnected ancilla circuit:
\begin{equation}
    \begin{quantikz}
    \lstick{$\lvert + \rangle$} & \gate{R_y(\theta)} & \meter{Z_{s_i'}}
    \end{quantikz}
\end{equation}
which is Eq.~\eqref{eq:Xmeascircuit}.

Deriving the circuits for the $\langle K_i\rangle_\theta$ measurements uses similar techniques, with two additions. The first is the insertion of Hadamards to invoke half-teleportation removals of measured qubits:
\begin{align}
\begin{quantikz}[align equals at=1.5]
    \lstick{} & \gate{Z_{s_{z, i}}} & \targ & \qw & \meter{Z_{s_i}}
    \\ \lstick{$\lvert + \rangle$} & \qw & \ctrl{-1} & \gate{R_y(\theta)} & \meter{Z_{s_{z, i}}}
\end{quantikz}&=
\begin{quantikz}[align equals at=1.5]
    \lstick{} & \gate{Z_{s_{z, i}}} & \gate{H^2} & \targ & \qw & \gate{H^2}  & \meter{Z_{s_i}}
    \\ \lstick{$\lvert + \rangle$} & \qw & \qw & \ctrl{-1} & \gate{R_y(\theta)} & \meter{Z_{s_{z, i}}}
\end{quantikz}
\\ &=
\begin{quantikz}[align equals at=1.5]
    \lstick{} & \gate{Z_{s_{z, i}}} & \gate{H} & \control{} & \qw & \meter{X_{s_i}}
    \\ \lstick{$\lvert + \rangle$} & \qw & \qw & \ctrl{-1} & \gate{R_y(\theta)} & \meter{Z_{s_{z, i}}}
\end{quantikz}
\end{align}

The additional subtlety is that the symmetry argument making the majority of the physical qubit measurements extraneous does not follow through as cleanly now that $Z$-measurements are added into the mix. Instead, we consider an additional step of tracing out the qubits of the chain not involved in the $K_i$ measurement. 

As a warm-up, we consider tracing out the first qubit of a bare cluster chain. Denote by $\rho^C_{1, \ldots, n}$ a cluster state with sites $1, \ldots, n$. Then:
\begin{equation}
    \text{Tr}_1(\rho^C_{1, \ldots, n}) = \bra{0}_1 \rho^C_{1, \ldots n} \ket{0}_1 + \bra{1}_1 \rho^C_{1, \ldots n} \ket{1}_1
\end{equation}
The first term corresponds to measuring $Z_1 = +1$ on the first qubit which occurs with probability $1/2$ and removes the qubit, leaving a $n-1$ site cluster chain:
\begin{equation}
    \bra{0}_1 \rho^C_{1, \ldots, n}\ket{0}_1 =  \frac{1}{2}\rho^C_{2, \ldots, n}
\end{equation}
to compute the other term, we pull out a cluster state stabilizer centered on the first qubit:
\begin{equation}
 \bra{1}_1 \rho^C_{1, \ldots, n}\ket{1}_1 =   \bra{1}_1 X_1Z_2\rho^C_{1, \ldots, n}X_1Z_2\ket{1}_1 = Z_2 \bra{0}_1 \rho^C_{1, \ldots, n}\ket{0}_1Z_2 = \frac{1}{2}Z_2\rho^C_{2, \ldots, n}Z_2
\end{equation}
Thus:
\begin{equation}
    \text{Tr}_1(\rho^C_{1, \ldots, n}) = \frac{1}{2}\rho^C_{2, \ldots, n} + \frac{1}{2}Z_2\rho^C_{2, \ldots, n}Z_2
\end{equation}
and the action of tracing out the first qubit is to have a $Z$ error on the second qubit of the cluster chain with probability one half. This argument can be applied repeatedly to trace out $m$ qubits, leaving a $n-m$ qubit cluster chain with a probabilistic $Z$ on the traced boundary.

Now, we consider the case of the dressed cluster chain, with the non-unitary $M_i(\theta)$ applied to each (non-boundary) site $i = 2, \ldots, n-2$. After tracing out the first qubit, we are left with:
\begin{equation}
    \rho_{2, \ldots, n} = \frac{1}{2}\left(\bigotimes_{i=2}^{n-1}M_i(\theta)\rho^C_{2, \ldots, n}\bigotimes_{i=2}^{n-1}M_i(\theta)\right) + \frac{1}{2}\left(\bigotimes_{i=2}^{n-2}M_i(\theta)Z_1\rho_{2, \ldots, n}^C Z_1\bigotimes_{i=2}^{n-2}M_i(\theta)\right)
\end{equation}
Now tracing out the second qubit in the $X$-basis, and using that $M_i(\theta)\ket{\pm}_i = (\cos(\frac{\theta}{2}) \pm \sin(\frac{\theta}{2}))\ket{\pm}_i$ and $M_i(\theta)Z_i = Z_iM_i(-\theta)$, we obtain:

\begin{align}
    \text{Tr}_2(\rho_{2, \ldots, n}) &= \bra{+}\frac{1}{2}\left(\bigotimes_{i=2}^{n-1}M_i(\theta)\rho^C_{2, \ldots, n}\bigotimes_{i=2}^{n-1}M_i(\theta)\right)\ket{+} + \bra{+}\frac{1}{2}\left(\bigotimes_{i=2}^{n-2}M_i(\theta)Z_1\rho_{2, \ldots, n}^C Z_1\bigotimes_{i=2}^{n-2}M_i(\theta)\right)\ket{+}
    \\ &+ \bra{-}\frac{1}{2}\left(\bigotimes_{i=2}^{n-1}M_i(\theta)\rho^C_{2, \ldots, n}\bigotimes_{i=2}^{n-1}M_i(\theta)\right)\ket{-} + \bra{-}\frac{1}{2}\left(\bigotimes_{i=2}^{n-2}M_i(\theta)Z_1\rho_{2, \ldots, n}^C Z_1\bigotimes_{i=2}^{n-2}M_i(\theta)\right)\ket{-}
    \\ &= \frac{(\cos(\frac{\theta}{2}) + \sin(\frac{\theta}{2}))^2}{2}\left(\bra{+}\bigotimes_{i=3}^{n-1}M_i(\theta)\rho^C_{2, \ldots, n}\bigotimes_{i=3}^{n-1}M_i(\theta)\ket{+} + \bra{-} \bigotimes_{i=3}^{n-1}M_i(\theta)\rho^C_{2, \ldots, n}\bigotimes_{i=3}^{n-1}M_i(\theta)\ket{-}\right)
    \\ &+ \frac{(\cos(\frac{\theta}{2}) - \sin(\frac{\theta}{2}))^2}{2}\left(\bra{-}\bigotimes_{i=3}^{n-1}M_i(\theta)\rho^C_{2, \ldots, n}\bigotimes_{i=3}^{n-1}M_i(\theta)\ket{-} + \bra{+} \bigotimes_{i=3}^{n-1}M_i(\theta)\rho^C_{2, \ldots, n}\bigotimes_{i=3}^{n-1}M_i(\theta)\ket{+}\right)
    \\ &= \bra{+}\bigotimes_{i=3}^{n-1}M_i(\theta)\rho^C_{2, \ldots, n}\bigotimes_{i=3}^{n-1}M_i(\theta)\ket{+} + \bra{-}\bigotimes_{i=3}^{n-1}M_i(\theta)\rho^C_{2, \ldots, n}\bigotimes_{i=3}^{n-1}M_i(\theta)\ket{-}
\end{align}
but the last expression simply corresponds to tracing out the second cluster state qubit without any $M_i(\theta)$ applied to it, so our argument from the warm-up applied, and this amounts to simply removing the qubit and placing a $Z$-error on qubit 3 with probability $\frac{1}{2}$:
\begin{equation}
    \text{Tr}_2(\rho_{2, \ldots, n}) = \frac{1}{2}\bigotimes_{i=3}^{n-1}M_i(\theta)\rho_{3, \ldots, n}^C \bigotimes_{i=3}^{n-1}M_i(\theta) + \frac{1}{2}\bigotimes_{i=3}^{n-1}M_i(\theta)Z_3\rho_{3, \ldots, n}^C Z_3\bigotimes_{i=3}^{n-1}M_i(\theta)
\end{equation}
the state is identical to our starting state (with the second qubit now removed and the $Z$-error shifted onto the third qubit), so as before, we may repeatedly apply the argument to trace out $m$ qubits of the dressed cluster state, which leaves a probabilistic $Z$ on the trace boundary.

Thus, to obtain the $\langle K_i\rangle_\theta$ circuits, we first trace out the majority of the state, leaving a probabilistic $Z$-error, and then proceed with half-teleportation and other similiar circuit identities to prune the circuits to the form of Eqs. \eqref{eq:Kboundmeascircuit}-\eqref{eq:Kbulkmeascircuit2}.

\section{Variational Anstaz}\label{app:ansatzphasetransition}
We again consider the interpolating Hamiltonian:
\begin{equation}
H(\alpha) = -\cos\alpha \sum_{i=1}^{N} K_i - \sin\alpha \sum_{i=2}^{N-1} X_i,
\end{equation}

with open-boundary stabilizers
\begin{equation}
K_1 = X_1 Z_2,\qquad  
K_i = Z_{i-1} X_i Z_{i+1}\ (2\le i\le N-1),\qquad  
K_N = Z_{N-1} X_N.
\end{equation}

A convenient non-unitary variational family is
\begin{equation}
|\psi(\theta)\rangle =
\Bigg[\prod_{i=2}^{N-1}\big(\cos\theta\, I + \sin\theta\, X_i\big)\Bigg]|C\rangle,
\qquad 0 \le \theta \le \frac{\pi}{4}.
\end{equation}
Because all factors commute and are Hermitian, and because the cluster stabilizers are mutually orthogonal, the normalization follows directly
\begin{equation}
\langle\psi(\theta)|\psi(\theta)\rangle
= \Big\langle C\Big|\prod_{i=2}^{N-1}\big(I+\sin(2\theta)X_i\big)\Big|C\Big\rangle = 1.
\end{equation}

Local expectation values are
\begin{align}
\langle X_i\rangle_\theta &= \sin(2\theta), && (2 \le i \le N-1),\\
\langle K_{1}\rangle_\theta = \langle K_{2}\rangle_\theta = \langle K_{N-1}\rangle_\theta = \langle K_{N}\rangle_\theta &= \cos(2\theta),\\
\langle K_i\rangle_\theta &= \cos^2\!\big(2\theta\big), && (3 \le i \le N-2).
\end{align}

For finite \(N\),
\begin{equation}
E(\theta;\alpha)
= -\cos\alpha\Big[4\cos(2\theta) + (N-4)\cos^2(2\theta)\Big]
   - \sin\alpha (N-2)\sin(2\theta).
\end{equation}

In the thermodynamic limit,
\begin{equation}
e(\theta;\alpha)
= -\cos\alpha\,\cos^2(2\theta) - \sin\alpha\,\sin(2\theta).
\end{equation}
Stationarity,
\begin{equation}
\partial_\theta e = 0
\quad\Longleftrightarrow\quad
\cos(2\theta)\,[2\cos\alpha\,\sin(2\theta) - \sin\alpha] = 0.
\end{equation}

The interior (unsaturated) solution exists when \(\tan\alpha \le 2\):
\begin{align}
\sin(2\theta^\ast) &= \frac{\sin\alpha}{2\cos\alpha} = \tfrac{1}{2}\tan\alpha,\\
\cos^2(2\theta^\ast) &= 1 - \frac{\tan^2\alpha}{4},\\
e^\ast(\alpha) &= -\cos\alpha - \frac{\sin^2\alpha}{4\cos\alpha}.
\end{align}

Beyond this range (\(\tan\alpha>2\)), the solution saturates:
\begin{align}
\theta^\ast &= \tfrac{\pi}{4},\\
\sin(2\theta^\ast) &= 1,\\
\cos(2\theta^\ast) &= 0,\\
e^\ast(\alpha) &= -\sin\alpha.
\end{align}

Hence the large-\(N\) variational ground state takes the form
\begin{equation}
|\psi_{\mathrm{gs}}(\alpha)\rangle =
\begin{cases}
\displaystyle
\Bigg[\prod_{i=2}^{N-1}\big(\cos\theta^\ast I+\sin\theta^\ast X_i\big)\Bigg]|C\rangle,
& 2\theta^\ast = \arcsin\!\Big(\frac{\sin\alpha}{2\cos\alpha}\Big),\quad \tan\alpha \le 2,\\[10pt]
\displaystyle
\Bigg[\prod_{i=2}^{N-1}\frac{I+X_i}{\sqrt2}\Bigg]|C\rangle,
& \tan\alpha > 2.
\end{cases}
\end{equation}

In the interior regime,
\begin{align}
\langle K\rangle &= \cos^2\!\big(2\theta^\ast\big) = 1 - \frac{\tan^2\alpha}{4},\\
\langle X\rangle &= \sin(2\theta^\ast) = \tfrac{1}{2}\tan\alpha.
\end{align}

Once saturation sets in,
\begin{align}
\langle K\rangle &= 0,\\
\langle X\rangle &= 1.
\end{align}

\section{Error metrics and Norms}\label{app:norm}

In this section we review the error definitions used throughout. Let the target logical unitary be $U_{\text{log}}$
and let the actually implemented quantum channel be $\mathcal{V}$.
The ideal channel associated with the target operation is  $\mathcal{U}(X)=U_{\text{log}}\,X\,U_{\text{log}}^{\dagger}.$

We define the error channel as the difference between the implemented and ideal channels,
\begin{align*}
\mathcal{E} := \mathcal{V}-\mathcal{U}.
\end{align*}
The corresponding error metric is defined as a scalar multiple of the Frobenius norm of the error channel,
\begin{align}\label{eq.Error}
\mathcal{D} := \sqrt{2}\,\|\mathcal{E}\|_{F},
\end{align}
where
\begin{align}
\|\mathcal{E}\|_{F}
:=\left(\sum_{\mu} \|\mathcal{E}(F_{\mu})\|_{2}^{2}
\right)^{1/2}.
\end{align}

Here $\{F_{\mu}\}$ is any operator basis satisfying
\begin{equation}
\operatorname{tr}(F_{\mu}^{\dagger}F_{\nu})=\delta_{\mu\nu},
\label{eq:orthobasis}
\end{equation}
and the Hilbert--Schmidt norm is given by
\begin{equation}
\|A\|_{2}^{2}=\operatorname{tr}(A^{\dagger}A).
\label{eq:HSnorm}
\end{equation}
For computational convenience, we choose $\{F_{\mu}\}$ to be the Pauli basis.

As an illustrative calculation, we estimate the error induced by a single
symmetry-breaking measurement at an odd site, which results in an imperfect
implementation of a logical \( Z \)-rotation.

The error channel itself is relevant only up to conjugation by unitaries, since
such conjugations preserve the error metric we employ. We therefore consider the
\emph{Pauli-basis representation of the error channel conjugated by the logical
\( Z \)-rotation}. Because the channel is trace-preserving and acts on a single
logical qubit, this representation is via a \( 3 \times 3 \) matrix.
Conjugation by a unitary corresponds to an orthogonal transformation in Pauli
space and hence leaves the Frobenius norm invariant.

Using the decomposition in Eq.~\eqref{Dephase} of the implemented channel
\( \mathcal{V} \), the Pauli-basis representation of the conjugated error channel
takes the form
\begin{equation}
\begin{pmatrix}
-2(\epsilon/4) & 0 & 0\\
0 & -2(\epsilon/4) & 0 \\
0 & 0 & 0
\end{pmatrix}.
\label{eq:PauliErrorMatrix}
\end{equation}

Its Frobenius norm evaluates to
\begin{equation}
\|\mathcal{E}\|_{F}
= \sqrt{(-\epsilon/2)^2 + (-\epsilon/2)^2}
= \frac{\epsilon}{\sqrt{2}}.
\label{eq:FrobeniusNorm}
\end{equation}

Accordingly, the error metric for a single-site rotation is
\begin{equation}
\mathcal{D} := \sqrt{2}\,\|\mathcal{E}\|_{F} = \epsilon .
\label{eq:ErrorMetric}
\end{equation}
This justifies our choice of the prefactor \(1/4\) in Eq.~\eqref{Dephase}.

\section{Device Error Parameters}\label{app:errors}
\begin{figure}[htbp!]
    \centering
    \begin{tikzpicture}[scale=0.7]
    \foreach \x in {0,...,13}{
        \node[circle, draw, thick] (\x) at (\x, 0) {\phantom{o}};
    };

    \node[circle, draw, thick] (14) at (0, -1) {\phantom{o}};
    \node[circle, draw, thick] (15) at (4, -1) {\phantom{o}};
    \node[circle, draw, thick] (16) at (8, -1) {\phantom{o}};
    \node[circle, draw, thick] (17) at (12, -1) {\phantom{o}};

    \foreach \x in {18, ..., 32}{
        \node[circle, draw, thick] (\x) at (\x-18, -2) {\phantom{o}};
    };

    \node[circle, draw, thick] (33) at (2, -3) {\phantom{o}};
    \node[circle, draw, thick] (34) at (6, -3) {\phantom{o}};
    \node[circle, draw, thick] (35) at (10, -3) {\phantom{o}};
    \node[circle, draw, thick] (36) at (14, -3) {\phantom{o}};

    \foreach \x in {37, ..., 51}{
        \node[circle, draw, thick] (\x) at (\x-37, -4) {\phantom{o}};
    };

    \node[circle, draw, thick] (52) at (0, -5) {\phantom{o}};
    \node[circle, draw, thick] (53) at (4, -5) {\phantom{o}};
    \node[circle, draw, thick] (54) at (8, -5) {\phantom{o}};
    \node[circle, draw, thick] (55) at (12, -5) {\phantom{o}};

    \foreach \x in {56, ..., 70}{
        \node[circle, draw, thick] (\x) at (\x-56, -6) {\phantom{o}};
    };

    \node[circle, draw, thick] (71) at (2, -7) {\phantom{o}};
    \node[circle, draw, thick] (72) at (6, -7) {\phantom{o}};
    \node[circle, draw, thick] (73) at (10, -7) {\phantom{o}};
    \node[circle, draw, thick] (74) at (14, -7) {\phantom{o}};

    \foreach \x in {75, ..., 89}{
        \node[circle, draw, thick] (\x) at (\x-75, -8) {\phantom{o}};
    };

    \node[circle, draw, thick] (90) at (0, -9) {\phantom{o}};
    \node[circle, draw, thick] (91) at (4, -9) {\phantom{o}};
    \node[circle, draw, thick] (92) at (8, -9) {\phantom{o}};
    \node[circle, draw, thick] (93) at (12, -9) {\phantom{o}};

    \foreach \x in {94, ..., 108}{
        \node[circle, draw, thick] (\x) at (\x-94, -10) {\phantom{o}};
    };

    \node[circle, draw, thick] (109) at (2, -11) {\phantom{o}};
    \node[circle, draw, thick] (110) at (6, -11) {\phantom{o}};
    \node[circle, draw, thick] (111) at (10, -11) {\phantom{o}};
    \node[circle, draw, thick] (112) at (14, -11) {\phantom{o}};

    \foreach \x in {113, ..., 126}{
        \node[circle, draw, thick] (\x) at (\x-112, -12) {\phantom{o}};
    };

    \draw[very thick] (0) -- (1) -- (2) -- (3) -- (4) -- (5) -- (6) -- (7) -- (8) -- (9) -- (10) -- (11) -- (12) -- (13);
    \draw[very thick] (18) -- (19) -- (20) -- (21) -- (22) -- (23) -- (24) -- (25) -- (26) -- (27) -- (28) -- (29) -- (30) -- (31) -- (32);
    \draw[very thick] (37) -- (38) -- (39) -- (40) -- (41) -- (42) -- (43) -- (44) -- (45) -- (46) -- (47) -- (48) -- (49) -- (50) -- (51);
    \draw[very thick] (56) -- (57) -- (58) -- (59) -- (60) -- (61) -- (62) -- (63) -- (64) -- (65) -- (66) -- (67) -- (68) -- (69) -- (70);
    \draw[very thick] (75) -- (76) -- (77) -- (78) -- (79) -- (80) -- (81) -- (82) -- (83) -- (84) -- (85) -- (86) -- (87) -- (88) -- (89);
    \draw[very thick] (94) -- (95) -- (96) -- (97) -- (98) -- (99) -- (100) -- (101) -- (102) -- (103) -- (104) -- (105) -- (106) -- (107) -- (108);
    \draw[very thick] (113) -- (114) -- (115) -- (116) -- (117) -- (118) -- (119) -- (120) -- (121) -- (122) -- (123) -- (124) -- (125) -- (126);

    \draw[very thick] (0) -- (14) -- (18);
    \draw[very thick] (4) -- (15) -- (22);
    \draw[very thick] (8) -- (16) -- (26);
    \draw[very thick] (12) -- (17) -- (30);

    \draw[very thick] (20) -- (33) -- (39);
    \draw[very thick] (24) -- (34) -- (43);
    \draw[very thick] (28) -- (35) -- (47);
    \draw[very thick] (32) -- (36) -- (51);

    \draw[very thick] (37) -- (52) -- (56);
    \draw[very thick] (41) -- (53) -- (60);
    \draw[very thick] (45) -- (54) -- (64);
    \draw[very thick] (49) -- (55) -- (68);

    \draw[very thick] (58) -- (71) -- (77);
    \draw[very thick] (62) -- (72) -- (81);
    \draw[very thick] (66) -- (73) -- (85);
    \draw[very thick] (70) -- (74) -- (89);

    \draw[very thick] (75) -- (90) -- (94);
    \draw[very thick] (79) -- (91) -- (98);
    \draw[very thick] (83) -- (92) -- (102);
    \draw[very thick] (87) -- (93) -- (106);

    \draw[very thick] (96) -- (109) -- (114);
    \draw[very thick] (100) -- (110) -- (118);
    \draw[very thick] (104) -- (111) -- (122);
    \draw[very thick] (108) -- (112) -- (126);
    \foreach \x in {0, ..., 126}{
        \node at (\x) {\small \x};
    };
\end{tikzpicture}
    \caption{Numbered layout of the \texttt{ibm\_quebec} chip.}
    \label{fig:ibmlayout}
\end{figure}
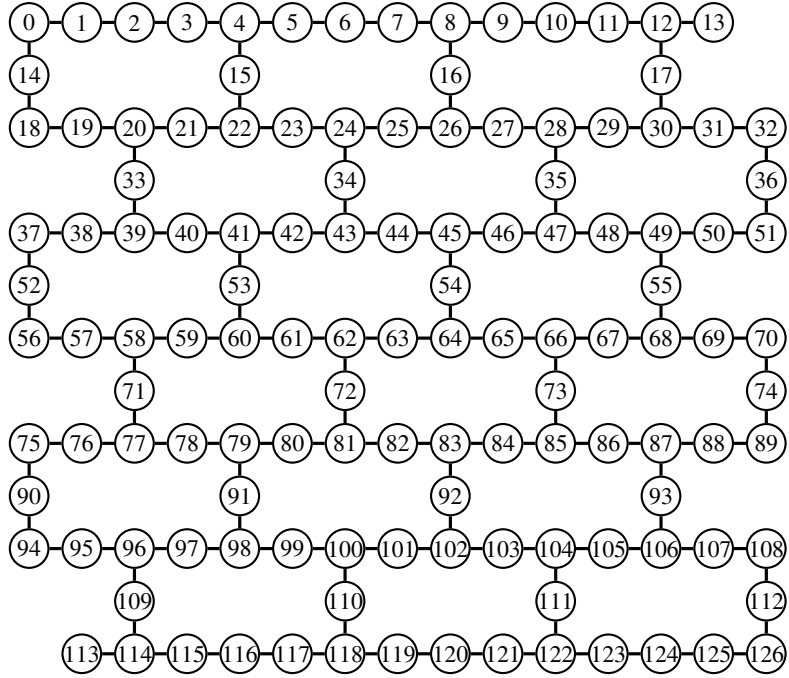

The 127-qubit \texttt{ibm\_quebec} device has the heavy-hex layout given in Fig.~\ref{fig:ibmlayout}. For each experiment we chose an appropriate subset of qubits with low reported error rates, which are provided in Tables \ref{tab:errors-VQE}-\ref{tab:errors-counterintuitive}.

Errors/gate times of $Z$-rotation gates are zero as these gates are performed virutally on the IBM hardware \cite{mckay_efficient_2017}.

\begin{table}[htbp!]
    \centering
    \begin{footnotesize}
    \begin{tabular}{|l|llllllllllllll|}
    \hline\hline
        Qubit & T1 ($\mu$s) & T2 ($\mu$s) & F (GHz) & A (GHz) & RE ($\times 10^{-3}$)  & M0P1 & M1P0 & RL (ns) & ID ($\times 10^{-4}$)  & SX ($\times 10^{-4}$) & X ($\times 10^{-4}$) & GT$_1$ (ns) & ECR ($\times 10^{-3}$)  & GT$_2$ (ns) \\
        \hline
        2 & 169.5 & 373.6 & 4.713 & -0.3127 & 0.24 & 0.0020 & 0.0028 & 835.6 & 0.9569 & 0.9569 & 0.9569 & 60.44 & & 
        \\ 64 & 201.0 & 195.1 & 4.885 & -0.3089 & 0.44 & 0.0054 & 0.0034 & 835.6 & 1.602 & 1.602 & 1.602 & 60.44 & & 
        \\ 105 & 348.4 & 105.0 & 4.813 & -0.3101 & 0.25 & 0.0038 & 0.0012 & 835.6 & 2.826 & 2.826 & 2.826 & 60.44 & & 
        \\ 12 & 456.8 & 346.9 & 4.708 & -0.3118 & 0.47 & 0.0046 & 0.0048 & 835.6 & 0.7994 & 0.7994 & 0.7994 & 60.44 & 13-12:2.912 & 593.8
        \\ 13 & 224.9 & 277.0 & 4.888 & -0.3088 & 0.45 & 0.0052 & 0.0038 & 835.6 & 1.544 & 1.544 & 1.544 & 60.44 & &
        \\ 27 & 429.2 & 377.2 & 4.759 & -0.3100 & 0.21 & 0.0028 & 0.0014 & 835.6 & 1.259 & 1.259 & 1.259 & 60.44 & 28-27:5.076 & 593.8
        \\ 28 & 201.5 & 40.47 & 4.848 & -0.3101 & 0.74 & 0.010 & 0.044 & 835.6 & 2.226 & 2.226 & 2.226 & 60.44 & & 
        \\ 40 & 350.2 & 92.10 & 4.882 & -0.3091 & 0.79 & 0.0056 & 0.010 & 835.6 & 1.311 & 1.311& 1.311& 60.44& 40-41:6.831 & 593.8
        \\ 41 & 145.4 & 242.6 & 4.795 & -0.3108 & 0.96 & 0.013 & 0.0064 & 835.6 & 2.019 & 2.019 & 2.019 & 60.44 & &
        \\ 116 & 224.6 & 247.6 & 5.007 & -0.3064 & 0.44 & 0.0054 & 0.0034 & 835.6 & 1.702& 1.702& 1.702& 60.44& 117-116:6.896 & 593.8
        \\ 117 & 283.5 & 271.0 & 4.873 & -0.3082 & 1.2 & 0.01 & 0.014 & 835.6 & 1.357 & 1.357 & 1.357 & 60.44 & &
    \\ \hline\hline
    \end{tabular}
    \end{footnotesize}
    \caption{Error parameters for involvled qubits in VQE experiment. T1 = T1 relaxtion time, T2 = T2 relaxation time, F = frequency, A = Anharmonicity, RE = Readout error, M0P1 = Probability of measuring $\ket{0}$ when preparing $\ket{1}$, M1P0 = Probability of measuring $\ket{1}$ when preparing $\ket{0}$, RL = Readout length, ID = Identity gate error, SX = $\sqrt{X}$ gate error, X = $X$ gate error, GT$_1$ = 1-qubit gate time, ECR = Echoed Cross Resonance gate error, GT$_2$ = 2-qubit gate time.} 
    \label{tab:errors-VQE}
\end{table}

\begin{table}[htbp!]
    \centering
    \begin{footnotesize}
    \begin{tabular}{|l|llllllllllllll|}
    \hline\hline
        Qubit & T1 ($\mu$s) & T2 ($\mu$s) & F (GHz) & A (GHz) & RE ($\times 10^{-2}$)  & M0P1 & M1P0 & RL (ns) & ID ($\times 10^{-4}$)  & SX ($\times 10^{-4}$) & X ($\times 10^{-4}$) & GT$_1$ (ns) & ECR ($\times 10^{-3}$)  & GT$_2$ (ns) \\
        \hline
        9 &  506.5 & 51.79 & 4.833 & -0.3104 & 1.3 & 0.020 & 0.0068 & 835.6 & 2.456 & 2.456 & 2.456 & 60.44 & 9-8:3.922 & 593.8
        \\ 8 & 471.0 & 80.50 & 4.744 & -0.3114 & 1.3 & 0.013 & 0.012 & 835.6 & 1.251 & 1.251 & 1.251 & 60.44 & 8-16:6.629 & 593.8
        \\ 16 & 265.1 & 85.29 & 4.900 & -0.3088 & 1.4 & 0.014 & 0.012 & 835.6& 1.405 & 1.405 & 1.405 & 60.44 & 16-26:6.843 & 593.8
        \\ 26 & 321.1 & 341.0 & 4.837 & -0.3108 & 0.41 & 0.0042 & 0.0040 & 835.6 & 3.726 & 3.726 & 3.726 & 60.44 & 26-25:6.751 & 593.8
        \\ 25 & 92.39 & 160.3 & 4.708 & -0.3115 & 0.67 & 0.010 & 0.003 & 835.6 & 1.523 & 1.523 & 1.523 & 60.44 & &
    \\ \hline\hline
    \end{tabular}
    \end{footnotesize}
    \caption{Error parameters for involved qubits in the 5-qubit COP experiment. T1 = T1 relaxtion time, T2 = T2 relaxation time, F = frequency, A = Anharmonicity, RE = Readout error, M0P1 = Probability of measuring $\ket{0}$ when preparing $\ket{1}$, M1P0 = Probability of measuring $\ket{1}$ when preparing $\ket{0}$, RL = Readout length, ID = Identity gate error, SX = $\sqrt{X}$ gate error, X = $X$ gate error, GT$_1$ = 1-qubit gate time, ECR = Echoed Cross Resonance gate error, GT$_2$ = 2-qubit gate time.}
    \label{tab:errors-COP}
\end{table}

\begin{table}[htbp!]
    \centering
    \begin{footnotesize}
    \begin{tabular}{|l|llllllllllllll|}
    \hline\hline
        Qubit & T1 ($\mu$s) & T2 ($\mu$s) & F (GHz) & A (GHz) & RE ($\times 10^{-2}$)  & M0P1 & M1P0 & RL (ns) & ID ($\times 10^{-4}$)  & SX ($\times 10^{-4}$) & X ($\times 10^{-4}$) & GT$_1$ (ns) & ECR ($\times 10^{-3}$)  & GT$_2$ (ns) \\
        \hline
        10 & 302.5 & 340.3 & 4.942 & -0.3083 & 0.46 & 0.0040 & 0.0052 & 835.6 & 1.378 & 1.378 & 1.378 & 60.44 & 11-10:9.597  & 593.8
        \\ 11 & 386.5 & 268.4 & 4.841 & -0.3094 & 3.8 & 0.036 & 0.039 & 835.6 & 1.391 & 1.391 & 1.391 & 60.44 & 11-12:3.777 & 593.8
        \\ 12 & 374.9 & 343.3 & 4.708 & -0.3118 & 0.60 & 0.0082 & 0.0038 & 835.6 & 1.008 & 1.008 & 1.008 & 60.44 & 17-12:11.72 & 593.8
        \\ 17 & 306.4 & 117.9 & 4.900 & -0.3088 & 0.73 & 0.0074 & 0.0072 & 835.6 & 1.881 & 1.881 & 1.881 & 60.44 & 30-17:8.009 & 593.8
        \\ 30 & 363.7 & 399.4 & 4.829 & -0.3099 & 0.81 & 0.0094 & 0.0068 & 835.6 & 1.449 & 1.449 & 1.449 & 60.44 & 31-30:5.012 & 593.8
        \\ 31 & 378.6 & 317.4  & 4.875 & -0.3094 & 1.8 & 0.020 & 0.016 & 835.6 & 1.334 & 1.334 & 1.334 & 60.44 & 31-32:5.240 & 593.8
        \\ 32 & 445.0 & 97.32 & 4.783 & -0.3100 & 1.2 & 0.014 & 0.010 & 835.6 & 2.340 & 2.340 & 2.340 & 60.44 & 36-32:6.255 & 593.8
        \\ 36 & 166.7 & 304.1 & 4.897 & -0.3083 & 1.4 & 0.012 & 0.016  & 835.6 & 2.858 & 2.858 & 2.858 & 60.44 & 51-36:5.492 & 593.8
        \\ 51 & 305.7 & 296.1 & 5.010 & -0.3067 & 1.0 & 0.009 & 0.012 & 835.6 & 1.584 & 1.584 & 1.584 & 60.44 &  & 593.8
    \\ \hline\hline
    \end{tabular}
    \end{footnotesize}
    \caption{Error parameters for involved qubits in the 9-qubit splitting experiment. T1 = T1 relaxtion time, T2 = T2 relaxation time, F = frequency, A = Anharmonicity, RE = Readout error, M0P1 = Probability of measuring $\ket{0}$ when preparing $\ket{1}$, M1P0 = Probability of measuring $\ket{1}$ when preparing $\ket{0}$, RL = Readout length, ID = Identity gate error, SX = $\sqrt{X}$ gate error, X = $X$ gate error, GT$_1$ = 1-qubit gate time, ECR = Echoed Cross Resonance gate error, GT$_2$ = 2-qubit gate time.}
    \label{tab:errors-splitting}
\end{table}

\begin{table}[htbp!]
    \centering
    \begin{footnotesize}
    \begin{tabular}{|l|llllllllllllll|}
    \hline\hline
        Qubit & T1 ($\mu$s) & T2 ($\mu$s) & F (GHz) & A (GHz) & RE ($\times 10^{-2}$)  & M0P1 & M1P0 & RL (ns) & ID ($\times 10^{-4}$)  & SX ($\times 10^{-4}$) & X ($\times 10^{-4}$) & GT$_1$ (ns) & ECR ($\times 10^{-3}$)  & GT$_2$ (ns) \\
        \hline
        64 & 174.4 & 205.0 & 4.885 & -0.3089 & 0.42 & 0.0052 & 0.0032 & 835.6 & 2.367 & 2.367 & 2.367 & 60.44 & 64-63:4.430 & 593.8
        \\ 63 & 300.91 & 397.5 & 4.993 & -0.3076 & 0.91 & 0.0076 & 0.011 & 835.6 & 2.523  & 2.523 & 2.523 & 60.44 & 62-63:9.335 & 593.8
        \\ 62 & 400.0 & 309.5 & 4.895 & -0.3094 & 0.94 & 0.010 & 0.0088 & 835.6 & 2.156 & 2.156 & 2.156 & 60.44 & 62-72:6.298 & 593.8
        \\ 72 & 280.5 & 76.23 & 4.856 & -0.3084 & 1.1 & 0.134 & 0.0086 & 835.6 & 1.567 & 1.567 & 1.567 & 60.44 & 81-72:3.212 & 593.8
        \\ 81 & 337.2 & 563.5 & 4.935 & -0.3085 & 1.4 & 0.016 & 0.011 & 835.6 & 1.054 & 1.054 & 1.054 & 60.44 & 80-81:5.441 & 593.8
        \\ 80 & 166.7 & 293.2 & 5.004 & -0.3076 & 2.0 & 0.021 & 0.019 & 835.6 & 2.623 & 2.623 & 2.623 & 60.44 & 80-79:6.969 & 593.8
        \\ 79 & 346.9 & 472.0 & 4.883 & -0.3091 & 1.4 & 0.019 & 0.010 & 835.6 & 1.540 & 1.540 & 1.540 & 60.44 & 91-79:7.036 & 593.8
        \\ 91 & 262.3 & 282.8 & 5.022 & -0.3073 & 0.77 & 0.0082 & 0.0072 & 835.6 & 2.338 & 2.338 & 2.338 & 60.44 & 91-98:4.876 & 593.8
        \\ 98 & 373.5 & 257.7 & 4.938 & -0.3079 & 2.6 & 0.013 & 0.038 & 835.6 & 2.464 & 2.464 & 2.464 & 60.44 & 97-98:5.159 & 593.8
        \\ 97 & 213.4 & 120.0 & 5.043 & -0.3067 & 0.94 & 0.010 & 0.0088 & 835.6 & 1.903 & 1.903 & 1.903 & 60.44 & 97-98:6.001 & 593.8
        \\ 96 & 275.17 & 229.3 & 4.998 & -0.3068 & 0.69 & 0.0072 & 0.0066 & 835.6 & 3.248 & 3.248 & 3.248 & 60.44 &  & 593.8
    \\ \hline\hline
    \end{tabular}
    \end{footnotesize}
    \caption{Error parameters for involved qubits in the 11-qubit counterintuitive regime experiment. T1 = T1 relaxtion time, T2 = T2 relaxation time, F = frequency, A = Anharmonicity, RE = Readout error, M0P1 = Probability of measuring $\ket{0}$ when preparing $\ket{1}$, M1P0 = Probability of measuring $\ket{1}$ when preparing $\ket{0}$, RL = Readout length, ID = Identity gate error, SX = $\sqrt{X}$ gate error, X = $X$ gate error, GT$_1$ = 1-qubit gate time, ECR = Echoed Cross Resonance gate error, GT$_2$ = 2-qubit gate time.}
    \label{tab:errors-counterintuitive}
\end{table}

\newpage
\section{Validity of the counterintuitive regime}\label{app:CIregimevalidity}
The theorem of \cite{adhikary_counter-intuitive_2023} is not applicable to our case in the sense that it is a theorem that applies in the thermodynamic limit. Indeed, in the limit of small chains it need not be the case that the densest packing of rotations is the most efficient. However below we show that these finite size effects vanish quickly with increasing chain lengths.

Concretely, let us consider the $XX$-rotated cluster state of Eq.~\eqref{eq:correlatedresource} and compute the loss in purity for different choices of rotation splitting. We use the exact expressions for the evolution of the logical 
Pauli operators under $z$-rotations (from Eqs. (11)/(12) of the supplemental material of \cite{adhikary_counter-intuitive_2023}):
\begin{equation}\label{eq:exactevo}
    \begin{bmatrix}
    \langle X \rangle
    \\ \langle Y \rangle
    \\ \langle Z \rangle
    \end{bmatrix}
    =
    \langle \left(\prod_{j \in \mathcal{R}}\begin{bmatrix} \cos\beta_j & -S_j\sin\beta_j & 0 \\ S_j\sin\beta_j & \cos\beta_j & 0 \\ 0 & 0 & 1 \end{bmatrix}\right)\rangle     \begin{bmatrix}
    \langle X \rangle_0
    \\ \langle Y \rangle_0
    \\ \langle Z \rangle_0
    \end{bmatrix}
\end{equation}
where $S_j = Z_jX_{j+1}I_{j+2}X_{j+3}\ldots X_{n-1}Z_n$ and the product is taken over all $Z$-rotation sites. For our choice of initial logical state of $\ket{+}$, we have $\langle X \rangle_0 = 1$ and $\langle Y\rangle_0 = \langle Z\rangle_0 = 0$.

For a given set of rotations, the matrix product of Eq.~\eqref{eq:exactevo} can be evaluated exactly. We can then evaluate the expectation values of products of string operators numerically to find the logical expectation values at the end of the protocol, and hence the loss in purity.

For the resource state of interest on the 11-qubit chain, there are 5 inequivalent ways of splitting up a total rotation by $\beta$:
\begin{itemize}
    \item Two rotations by $\beta/2$, spaced by $\Delta = 2$ sites.
    \item Two rotations by $\beta/2$, spaced by $\Delta = 4/6$ sites (this yields the same result as the string order parameter converges to a constant beyond 4 sites).
    \item Three rotations by $\beta/3$, each spaced by $\Delta = 2$ sites.
    \item Three rotations by $\beta/3$, the first two spaced by $\Delta = 4$ sites, and the next spaced by $\Delta = 2$ sites.
    \item Four rotations by $\beta/4$, each spaced by $\Delta = 2$ sites.
\end{itemize}

\begin{figure}[htbp!]
    \centering
    \includegraphics[width=0.5\linewidth]{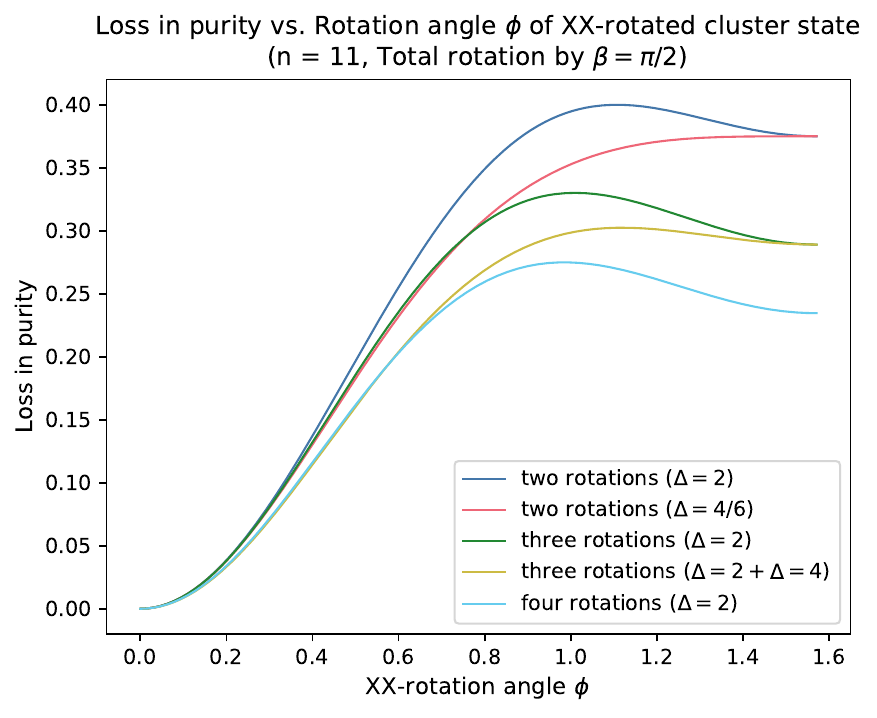}
    \caption{Comparison between different splitting schemes of the logical loss in purity arising from a total rotation of $\beta = \pi/2$ for an 11-qubit chain. The ordering of the curves is dependent on the $XX$-rotation angle $\phi$ applied to the cluster state.}
    \label{fig:11qubitCIanalysis}
\end{figure}

The loss in purity in each of those cases for a range of $\phi$ is provided in Fig.~\ref{fig:11qubitCIanalysis}, for a total maximally-symmetry breaking rotation angle of $\beta = \pi/2$. The theorem not applying manifests in the observation that the densest packing (four rotations spaced by $\Delta = 2$ sites, Red) is not the optimal protocol for $\phi < \phi_c \approx 0.574$, wherein it is outperformed by three rotations (red).

\begin{figure}[htbp!]
    \centering
    \includegraphics[width=0.5\linewidth]{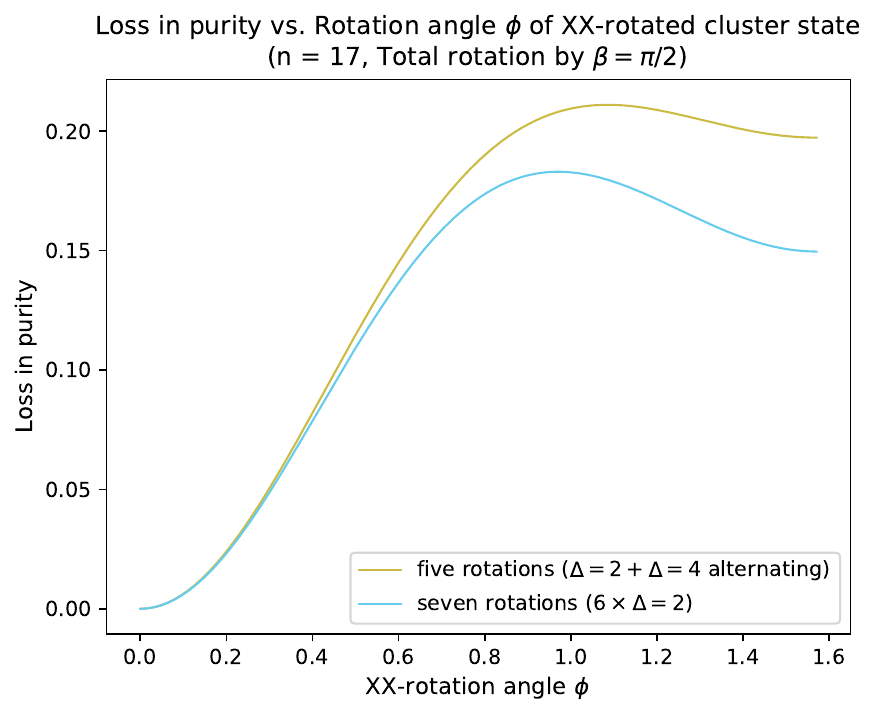}
    \caption{Comparison between alternating splitting distances and densest packing of splitting for the logical loss in purity arising from a total rotation of $\beta = \pi/2$ for an 17-qubit chain. The densest packing is optimal for all $XX$-rotation angles.}
    \label{fig:17qubitCIanalysis}
\end{figure}

Note that our experimental results indeed work in the regime with $\phi > \phi_c$ where the densest packing is indeed optimal. In addition, we can demonstrate that the finite size effect of having a state-dependent optimal protocol vanishes as we scale up the system size. Scaling the chain up to 17 qubits, we can tile/double the two competing protocols and compare five rotations by $\beta/5$ (spaced by $\Delta = 4$ and $\Delta = 2$ in alternating fashion) and seven rotations by $\beta/7$ (spaced uniformly by $\Delta = 2$). Doing so, we obtain Fig.~\ref{fig:17qubitCIanalysis}, wherein the crossover point $\phi_c$ is seen to go to zero, and the densest packing of rotations is indeed the most optimal for angles $\phi$.

\end{appendices}

\end{document}